\documentclass[10pt,journal,compsoc]{IEEEtran}

\usepackage{graphicx}
\usepackage{color}
\usepackage{multirow}   
\usepackage{hyperref}
\usepackage{epstopdf}
\usepackage[normalem]{ulem}
\usepackage{url}
\usepackage{verbatim} %
\usepackage{xspace} 
\usepackage{paralist} 
\usepackage{array}

\usepackage{ifthen}
\newboolean{longver}
\setboolean{longver}{true}%

\newcommand{\system}[1]{{\tt SkillVet}\xspace}

\newif\ifcomment
\commenttrue %
\ifcomment
\newcommand{\guillermo}[1]{{\bf \textcolor{blue}{GST: #1}}}
\newcommand{\jose}[1]{{\bf \textcolor{red}{JMS: #1}}}
\newcommand{\jide}[1]{{\bf \textcolor{red}{JSE: #1}}}
\newcommand{\xavi}[1]{{\bf \textcolor{green}{X: #1}}}
\newcommand{\old}[1]{{\textcolor{red}{\st{#1}}}}
\newcommand{\future}[1]{{\textcolor{green}{#1}}}
\else
\newcommand{\guillermo}[1]{}
\newcommand{\jose}[1]{}
\newcommand{\jide}[1]{}
\newcommand{\xavi}[1]{}
\newcommand{\old}[1]{}
\newcommand{\future}[1]{}
\fi

\newcommand{\done}[1]{}

\newif\ifreview
\reviewtrue
\ifreview

\else

\fi

\ifCLASSOPTIONcompsoc
  \usepackage[nocompress]{cite}
\else
  \usepackage{cite}
\fi

\ifCLASSINFOpdf
\else
\fi

\begin{document}

\title{SkillVet: Automated Traceability Analysis of Amazon Alexa Skills}

\author{Jide~S~Edu,
        Xavier~Ferrer-Aran,
        Jose~Such,
        Guillermo~Suarez-Tangil%
\IEEEcompsocitemizethanks{\IEEEcompsocthanksitem J. Edu, X.Ferrer~Aran, J.Such  were with  King's College (KCL) London. J.Such was also with Universitat Polit\`ecnica de Val\`encia.\protect\\
E-mail: \{jide.edu, xavier.ferrer\_aran,jose.such\}@kcl.ac.uk.
\IEEEcompsocthanksitem G.Suarez-Tangil was with IMDEA Networks Institute and KCL.\protect\\
E-mail: guillermo.suarez-tangil@ imdea.org}
}

\ifthenelse{\boolean{longver}}{}{\markboth{IEEE TRANSACTIONS ON DEPENDABLE AND SECURE COMPUTING VOL. XX, NO. XX, XXXX 2021}%
{XX \MakeLowercase{\textit{et al.}}: SkillVet: Automated Traceability Analysis of Amazon Alexa Skills}}

\IEEEtitleabstractindextext{%
\begin{abstract}
Skills, are essential components in Smart Personal Assistants (SPA). 
The number of skills has grown rapidly, dominated by a changing environment that has no clear business model. 
Skills can access personal information and this may pose a risk to users. 
However, there is little information about how this ecosystem works, let alone the tools that can facilitate its study. 
In this paper, we present the largest systematic measurement of the Amazon Alexa skill ecosystem to date. %
We study developers' practices in this ecosystem, including how they collect and justify the need for sensitive information, by designing a methodology to identify over-privileged skills with broken privacy policies. 
We collect 199,295 Alexa skills and uncover 
that around 43\% of the skills (and 50\% of the developers) that request these permissions follow bad privacy practices, including (partially) broken data permissions traceability. 
In order to perform this kind of analysis at scale, we present \system{} that leverages machine learning and natural language processing techniques, and generates high-accuracy prediction sets.
We report several concerning practices, including how developers can bypass Alexa's permission system through account linking and conversational skills, and offer recommendations on how to improve transparency, privacy and security.
Resulting from the responsible disclosure we did, 13\% of the reported issues no longer pose a threat at submission time.
\end{abstract}
}

\maketitle

\IEEEdisplaynontitleabstractindextext

\IEEEpeerreviewmaketitle

\section{Introduction}
\label{Sec:Introduction}

Smart Personal Assistants (SPA) are changing how users interact with technology and the way they consume services~\cite{ammari2019music}, driving new experiences and expectations~\cite{Noura_Abdi_2019,abdi2021privacy}. Advances in natural language processing enable SPA %
to take voice commands, which they process using machine learning to deduce users' intentions and fulfill users' requests. SPA are becoming very popular. 
According to recent reports, nearly 90 million adults in the US use at least one SPA device~\cite{SPA_adoption}. 
The adoption is expected to rocket even further as SPA integrate %
into smartphones~\cite{alexa_smartphones} or as a chatbot~\cite{alexa_chatbot}.

Despite their growing popularity, intrinsic features of SPA expose users to various risks~\cite{DBLP:journals/corr/abs-1903-05593,abdullah2020sok}, including 1) the open nature of the voice channel, 2) the complexity of their architecture, and 3) the use of a wide range of underlying technologies. %
Although considerable research has been recently conducted to understand and address these risks, many of them remain unresolved~\cite{DBLP:journals/corr/abs-1903-05593}. In particular, much of the research has focused on architectural elements of SPA devices (e.g., Amazon Echo)~\cite{Lei_Xinyu,kepuska_bohouta_2018,feng_fawaz_shin_2017} and the speech recognition capabilities of SPA~\cite{abdullah2020sok,brasser_frassetto_riedhammer_sadeghi_schneider_weinert_2018}.
However, SPA incorporate voice-driven applications generally developed by third parties, so-called \emph{skills}. 
The skills ecosystem widens the attack surface of SPA~\cite{DBLP:journals/corr/abs-1903-05593}, as malicious third-party actors may develop potentially harmful or unwanted software~\cite{Security_Research_Labs, 217575}.
Unfortunately, it is unclear what are the risks third-party skills may pose to users, beyond the potential for skill impersonation~\cite{zhang2019dangerous,217575} and other potentially suspicious behaviors~\cite{255322}. 
Authors in~\cite{chengdangerous2020} have very recently shown that the Amazon skill certification process can be subverted by crafting mock policy-violating skills. 
However, there is a critical knowledge gap in understanding what are the \emph{actual} data collection practices of skill developers in the wild.  
This gap prompts four core research questions: 
1) {\it What is the current state of affairs in the third-party ecosystem of skills?} 
2) {\it Is the collection of personal information explained?}
3) {\it Can we pinpoint fine-grained privacy issues (i.e., at the permission level)?} 4) {\it How can we do this in an automated way at scale?}

In this paper, we perform the first systematic measurement study of Amazon Alexa skills across  {the entire} marketplace of 11 different countries to shed light on the third-party skills ecosystem and developers' practices. 
We characterize skills using their market categories, names, and developers, among others. We also characterize the data permissions they request during runtime and their traceability with data practice statements in the skill's privacy policy. 
Since it is time-consuming to analyze privacy policies manually, we also propose an automated machine-learning-based system, \system{}, to help with this at scale. 
\system{} identifies relevant policy statements and assesses the traceability as complete, partial, or broken. Our system overcomes challenges in modeling privacy policies, such as contradictions like negations and statements related to multiple types of personal information. 

We then analyze suspicious skills in the look for concerning privacy practices. We do this by systematically analyzing the traceability between the permissions requested by the skill and the data practices stated by the skill developer in the privacy policy. 
This is done following a black-box approach since the skills' code or executable is not available --- skills \emph{run in the cloud instead of the users' device}%

Our study makes the following key contributions:

\begin{compactenum}
    
    \item We perform the
    largest systematic study of the Alexa skill ecosystem across 11 Amazon marketplaces (\S\ref{sec:overview}). 
    We take a snapshot of all markets and measure the distribution of skills, developers' diversity, and the invocation names used for {199k skills.}
    
    \item We study the permission model Amazon offers to developers, and investigate how skills make use of permissions to collect personal data (\S\ref{sec:permissions}). 
    We then design a methodology that systematically identifies potential privacy issues by analyzing the traceability between the permissions and the data practices stated by developers. We find a wealth of over-privileged skills with \emph{broken} traceability.%
    
    \item We design a system to automate the traceability analysis at scale, \system{}, based on machine learning and natural language processing (\S\ref{sec:system}). 
    Our system can be deployed by SPA providers to automatically conduct traceability analysis across marketplaces regularly and by developers to check the traceability of their skills. 
    We demonstrate \system{}'s accurate performance. 
    
    \item We provide insights regarding the developers' business models and concerning practices we observe (\S\ref{sec:bypassing}). 
    These practices include skills bypassing the permission system of Alexa to collect personal data from users by: a) account linking, and b)  conversational skills. 
    We provide evidence of skills (and their traceability) conducting these practices and discuss the implications of our findings (\S\ref{sec:DISCUSSION}).
    
\end{compactenum}

\noindent To foster research in the area, we release our datasets together with the ground-truth we have produced and our code in {\url{https://github.com/jideedu/SkillVet}} %

\section{The Skill Ecosystem}
\label{sec:overview}

\ifthenelse{\boolean{longver}}{In this section, we present our characterization of all Alexa skills across 11 markets.

\subsection{SPA Architecture \& Skills}
\label{sec:Background}
SPA devices such as Amazon Echo are equipped with a microphone and a voice interpreter that records utterances. Voice interpreters are often pre-activated and run in the background, where they wait for the users to say the wake-up word~\cite{2800-18}. Once they identify the wake-up word, the SPA device puts itself into recording mode. In recording mode,  utterances are sent to the SPA cloud~\cite{Understand_Skills}, where they are processed using Machine Learning (ML) and Natural Language Processing (NLP) to deduce the user's intention. Once the intention is identified, the SPA delegates the request to the most suitable voice application --- called \emph{skill} in Amazon Alexa --- that matches the user intention. The answers or information generated by the skill, as a response to the user request, will then be given back to the users by the SPA.}

Skills are, in a way, similar to mobile apps, as they enhance the capability of the ecosystem they are part of. They extend SPA functionality, allowing them to offer extensive services tailored to users' interests and experiences. The SPA skill ecosystem provides an environment that enables the user to run functions such as music playback, online purchase, appointment booking, and different innovative home automation tasks. For instance, a user can use a ``Uber'' skill to order a ride through the Amazon Echo.
\ifthenelse{\boolean{longver}}{
Unlike mobile apps, however, skills do not run on any user-controlled device. 
Instead, skills either run in the Amazon cloud (e.g. as an AWS Lambda function) or in a server controlled by the developer \cite{Alexa_CustomWeb}.
Therefore, the source code of skills is not available and instrumentation beyond voice interaction extremely difficult.}

A skill can simply be activated with an invocation utterance, i.e.: the trigger phrase and the skill invocation name. For example, to invoke the ``My Chef'' skill, a user can say \emph{`Alexa, open My Chef'}, or \emph{`Alexa, start My Chef'}. 
Note that ``Alexa, open'' and ``Alexa, start'' are trigger phrases to put the SPA into recording mode. When a skill is successfully invoked, specific phrases can then be uttered to call intents (functions) within the skill. However, in some cases, a skill needs to be enabled before it can be invoked, for instance, when the skill requires permissions. 
Skills can be enabled via compatible apps or online at the skill store. 

\subsection{Threat Model}
\label{sec:threat}

The attack surface of SPA is considerable large and can lead to several security and privacy issues --- for an extensive review of the literature refer to  \cite{DBLP:journals/corr/abs-1903-05593}. One of the attack entry points in SPA, and the focus of this article, is SPA skills. 
Unlike mobile apps, skills do not run on any user-controlled device. 
Instead, skills either run in the Amazon cloud (e.g., as an AWS Lambda function) or in a server controlled by the developer \cite{Alexa_CustomWeb}. %
This could allow a malicious developer to sneak malicious code into their software via the application backend~\cite{Security_Research_Labs}. %
They could manipulate skills to covertly introduce misperception about reported events \cite{SHAREVSKI2021102604} and also use skills with duplicate invocation names, similar phonemes, or paraphrased invocation names to impersonate another skill \cite{zhang2019dangerous,217575}. 
Recently, there has been report of how malicious actors, through a phishing attack, could steal a token that can let them add suspicious skills to the users’ account and access voice history~\cite{Amazon_security_bug}.

In this paper, we also consider the threat from skills developed by third-party developers, but focus on the data practices of such developers, whether malicious or just negligent. %
Skills allow users to interact with their services and require the exchange of data.
This could be data that are part of the user's account with Amazon, such as users~\emph{location, mobile number, email address, name, address}~\cite{Alexa_permissions_list}, sensitive data inferred from user actions with the skills, or personal data collected by the skills during their conversations with the users.
By considering skill developers as adversaries, we look into potentially over-privileged skills to understand what personal information they collect. This issue has been explored by several previous studies on permission gaps in other domains \cite{6494934, 7417621} but not in SPA and their skills. %
While declaring more permissions than required seems to have no impact on a skill's functionality, it could, however, be leveraged by a malicious developer to achieve malicious goals \cite{6494934}, which could impact the users' privacy.

This research presents \system{}, a system that could detect skills with inappropriate usage of data permissions at large and help protect users' privacy. For instance, ``Mock Interview'' skill \emph{by Graylogic Technologies} as detailed later. In addition, our system could help users understand skills data practices to make a better decision about their data.

\ifthenelse{\boolean{longver}}{
\subsection{Crawling Amazon Alexa}
\label{sec:crawl_pages}

Amazon lists the top most popular skills organized by category (and sub-category) up until a maximum of 400 pages per category (23 categories and 66 subcategories in total).
However, the skill marketplace is not entirely listed in the Amazon Alexa index.
Thus, crawling this index alone does not make the data collection exhaustive. 

We overcome this limitation by %
looking into sub-categories for each category per marketplace. 
Every subcategory across marketplaces returns less than 400 pages, so we can crawl all the skills within them. There are only two subcategories with more than 400 pages, both in the US (out of a total of 66 subcategories): Knowledge \& Trivia (from the Games \& Trivia category), and Education \& Reference (from the same category). 
For these two specific cases, we then use a best effort approach and re-crawl skills in those subcategories with a different sorting of the skill listing (i.e., ordering by {\it average costumer review} in addition to the default one --- {\it featured}). 
Thus, we are confident that we capture most, if not all, of the market space, as we collect all the skills available across all 11 marketplaces but the US. 
In the US, we collect more skills than the sum of all approximate numbers Amazon gives per subcategory,\footnote{Note that Amazon does not give the exact number of skills per category/subcategory when they contain 1,000 skills or more, but it rounds it down to the closest number in units of thousands.} which is $\approx$57,000 at crawling time (see below for the date), and we are able to crawl 61,362 skills from the US market, as detailed later. %
} 

\subsection{Collection and Characterization}

Amazon uses different online marketplaces to offer targeted services to several market segments. 
Amazon has 14 online marketplaces, out of which 11 have an online store with third-party skills. 
We crawl these 11 markets: US, UK, India, Australia, Canada, Germany, Japan, Italy, Spain, France, and Mexico. 
Skills are only available to registered users and a user can only be linked with one marketplace at a time. 

For all the skills we collect, we make a characterization of the market category they belong to, the utterances that activate the skill, and the developer that has created it. Two further key elements we extract are i) the permissions that the skill requires to access personal information from the Amazon user's account \textbf{via the Alexa API at runtime}, which Amazon makes publicly available in the marketplace; and ii) the privacy policy declared by the developers. 
We then use these two elements, together with an interactive analysis, to study when there is a privacy violation or a concerning practice that may harm users.

Unlike other platforms like Android, Amazon Alexa runs the skills in the cloud and \emph{the code of the skills is not publicly available}. 
To analyze the ecosystem of skills, we built a Web scrapper that recursively crawls all available marketplaces and extracts the metadata published by Amazon about the skills. 
Our crawler exhaustively collects all skills available in the Amazon Alexa marketplaces (for details refer to \ifthenelse{\boolean{longver}}{Appendix~\ref{sec:crawl_pages}}{\cite{edu2021skillvet}}).
Note that one skill can belong to different categories and be hosted in multiple marketplaces.
The URL may have different parameters in these cases, but the path displays a unique identifier per skill. 
We identify unique skills posted across markets in a process we call de-duplication. 

We use the following attributes (which we scrape from the marketplace) to characterize every skill: invocation name, permissions, the category, developer's information, privacy policy, terms of use, skill description, rating information and reviews. 
While new skills are added daily, we base our study on those skills accessible to users between the 15th of July 2020 and the 29th of July 2020.
This is the time-frame that took our crawler to visit all marketplaces. 
\ifthenelse{\boolean{longver}}{Since our data collection requires issuing a request to web services, we avoid generating unnecessary traffic that may affect their normal operations and limit the rate at which we generate our requests.}

\subsection{Skills and Developers}
\label{subsec:skills-developers}

Table~\ref{tb:Skills and Developers} shows the breakdown of the total number of Alexa skills and developers across Amazon marketplaces. 
Overall, we see {199,295 skills published by 88,391 developers. }
Note that most of these skills have been developed in just a few years.\footnote{Alexa had only 135 skills in early 2016~\cite{Amazon_skills_2016}.} 
English-speaking marketplaces host the highest number of skills and developers. 
On the other end, the smallest market is Mexico with {1,972 skills and representing 1\% of our dataset.} 
However, there is overlap across markets as developers can publish the same skills in different markets~\cite{Alexa_multi_language}. %
This overlap is larger among English-speaking markets such as the US, UK, CA, and AU where Amazon tends to migrate existing skills~\cite{alexa_English_Variant}. 
In particular, {we observe that close to 20,000 different skills in the UK market are also in the US market. 
When we de-duplicate skills, we observe that 53.76\% (47,520) of the developers and 56.2\% (112,029) of the skills we collect are unique. 
In what follows we look at unique skills unless we analyze specific markets.} 

\begin{table}[th]
\renewcommand{\arraystretch}{1.0}
\centering
\small{
\caption{Number of skills \& developers. English-speaking markets represent 82.74\% (92k skills and 36k devs).}
\label{tb:Skills and Developers}
\scalebox{0.9}{
\begin{tabular}{l|cc|cc}
\multicolumn{1}{c}{\multirow{2}{*}{Marketplace}} & \multicolumn{2}{c}{Skills}  & \multicolumn{2}{c}{Developers}  \\
\multicolumn{1}{c}{}                             & N       & Percent           & N      & Percent                \\ 
\hline
US                                               & 61,362  & 30.79\%           & 27,030 & 30.58\%                \\
UK                                               & 32,822  & 16.47\%           & 14,487 & 16.39\%                \\
India                                            & 29,344  & 14.72\%           & 12,823 & 14.51\%                \\
Canada                                           & 26,027  & 13.06\%           & 11,509 & 13.02\%                \\
Australia                                        & 23,909  & 12.00\%           & 10,854 & 12.28\%                \\
Germany                                          & 9,096   & 4.56\%            & 3,385  & 3.83\%                 \\
Spain                                            & 4,759   & 2.39\%            & 2,441  & 2.76\%                 \\
Italy                                            & 4,203   & 2.11\%            & 2,049  & 2.37\%                 \\
Japan                                            & 3,513   & 1.76\%            & 1,407  & 1.59\%                 \\
France                                           & 2,288   & 1.15\%            & 1,194  & 1.35\%                 \\
Mexico                                           & 1,972   & 0.99\%            & 1,212  & 1.37\%                 \\ 
\hline
\multicolumn{1}{r|}{Total}                       & 199,295 & 100.00\%          & 88,391 & 100.00\%               \\
\multicolumn{1}{r|}{Unique}                      & 112,029 & \textit{ 56.2}\%  & 47,520 & \textit{ 53.76}\%      \\
\hline
\end{tabular}}}
\end{table}

\begin{figure}[th]
    \centering
    \vspace{-3pt}
    \includegraphics[width=0.75\columnwidth, trim=1 4 148 60, clip]{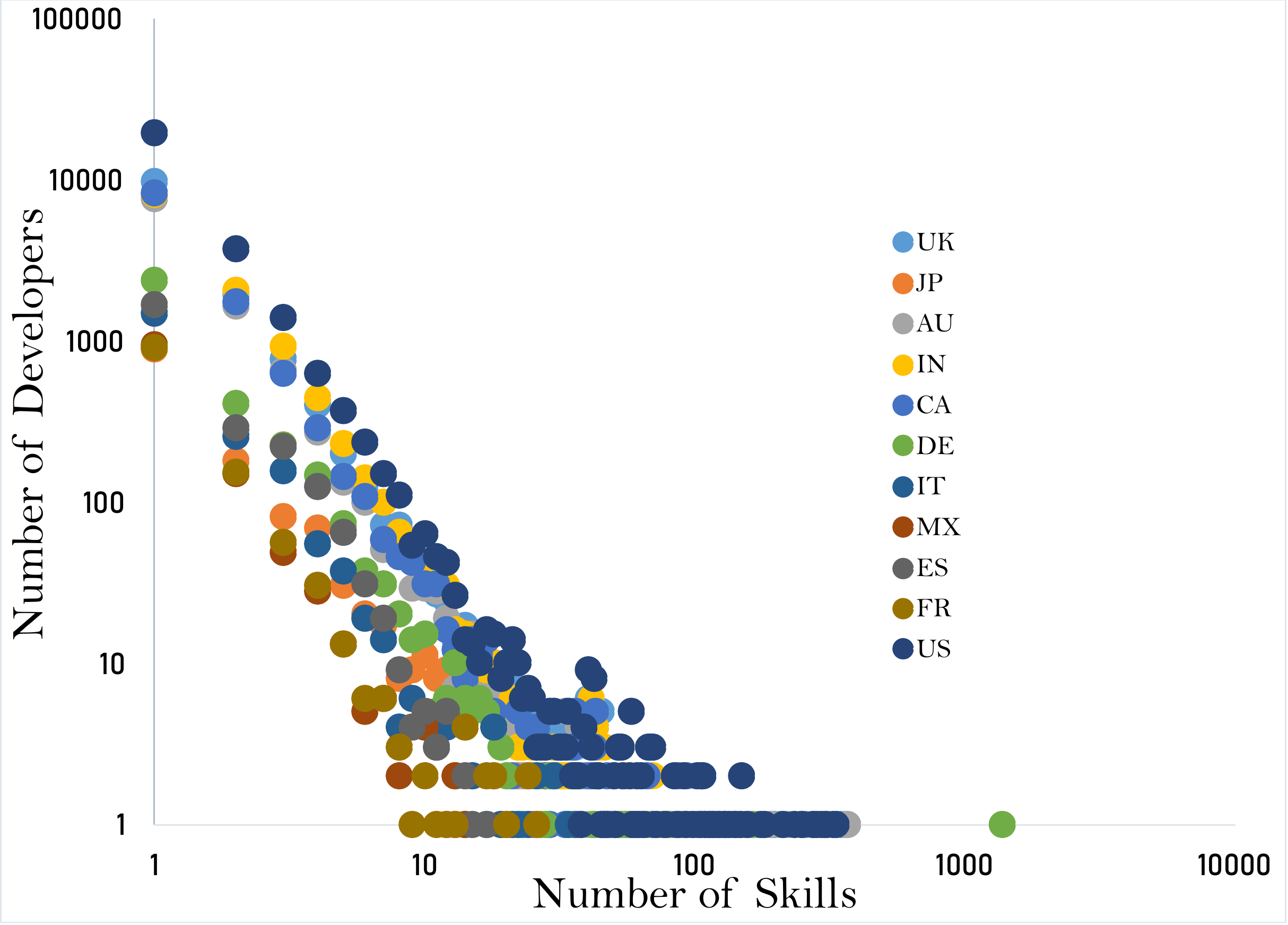}
        \vspace{-7pt}
    \caption{
    Developers vs the number of skills they publish per marketplace.} 
    \label{fig:developers_to_skills.png}
    \vspace{-2.5mm}
\end{figure}

\noindent\textbf{Developers.} We illustrate the relationship between developers and skills per marketplace in Figure~\ref{fig:developers_to_skills.png}. 
While most of the developers publish only one skill, a few of them have tens, and a handful have hundreds and even thousands of skills. 
{In particular, 125 developers have published more than 50 skills, 69 developers more than 100 skills, and 5 developers more than 1K skills.}
We then study the description of all the skills published by prominent developers and see that many implicitly indicate that the skill has been developed with an automated platform. 
This is usually done by referring to the name or the URL of the framework that produced the skill. 
Popular among them are \emph{VoiceApps.com}, \emph{VoiceXP.com} and \emph{VoiceFlow.com}.
These platforms provide free sample skills that a developer can customize regardless of their technical background. 
In particular, they offer visual, drag-and-drop editors that %
lowers down any development barriers. 
We mine all descriptions in our dataset and see that at least 7\% of the skills have been developed with these platforms. 
In particular, we observe a ratio of 1 developer to 30 skills with the most popular automated platform (VoiceApps.com).

\ifthenelse{\boolean{longver}}{}{
\noindent\textbf{Skill Invocation Names.}
Previous research showed that an attacker could impersonate skills by crafting specific malicious skill invocation names~\cite{zhang2019dangerous,217575, chengdangerous2020} by reusing the same as or (phonetically) similar invocation names to other skills. 
We measure the prevalence of same and similar skill invocation names across the Alexa ecosystem to understand the potential for this risk.
We find 24,468 unique skills (21.8\%), 6,876 (11.2\%) in the US, with the same invocation name as another/other skill/s after cross-market de-duplication.
We also find a considerable amount of very similar skill invocation names. %
For instance, we see that between 5\% and 15\% of all the skills in the US have a different but phonetically similar name. This is in line with recent measurements in less markets \cite{lentzschhey}. %
For detailed results, including the most popular names, the breakdown per country, 
the method to measure phonetic similarity, and the breakdown per similarity, refer to \cite{edu2021skillvet}. 
}

\ifthenelse{\boolean{longver}}{
\subsection{Invocation Name Reuse}
\label{sec:appendix:namereuse}

{There are 99,521 skill invocation names out of all 112,029 unique skills in our dataset.}
Figure~\ref{fig:skillnames to Developers} shows a scatter plot with the aggregate of invocation names per marketplace. 
In particular, the figure shows the total number of skill invocation names ($Y$ axis) w.r.t. how many other skills have the same invocation name ($X$ axis). 
We see that the number of skills with unique invocation names ($X = 0$) ranges from {1,914 (Mexico) to 54,486 (US) and totals 66,087} after cross-market skill de-duplication.
Recall that we only consider unique skills when aggregating results across marketplaces.
When we look at $X > 0$, we see 24,468 skills with invocation name reuse ({10,480 developers reuse 24,468 skill invocation names}). 
As values of $X$ increase, we see more popular names. 
An important number of skills share the same invocation name in different markets. 
We observe that India predominantly appears in cases with large values of $X$, %
with about 34\% of their invocation names being reused.

\begin{figure}[ht]
    \centering
    \includegraphics[width=.75\columnwidth, trim=9 8 10 65, clip]{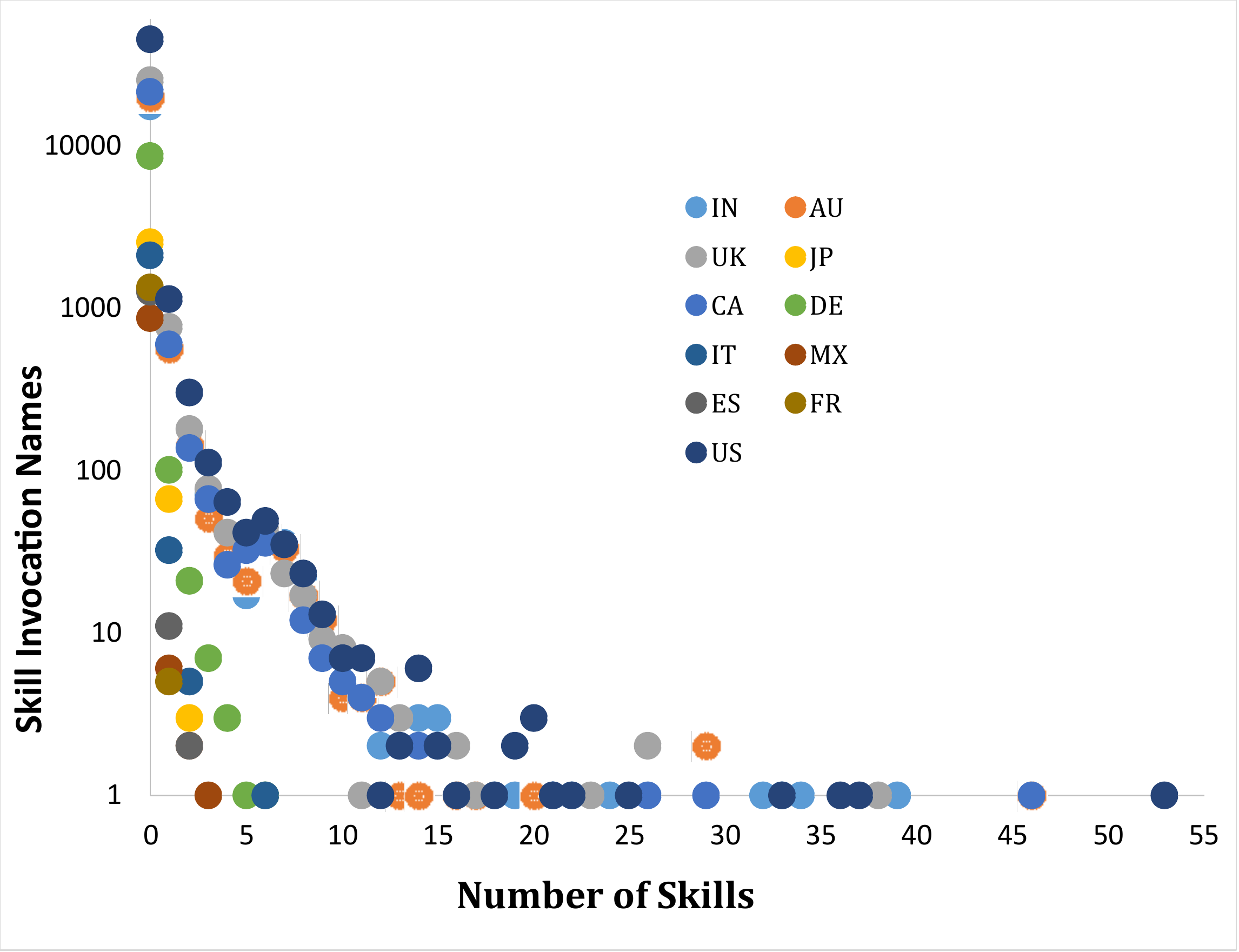}
    \vspace{-7pt}
    \caption{Number of skills with the same name across Alexa marketplaces. Most of the skills use a unique invocation name, but a high proportion of skills use the same invocation name.}
    \label{fig:skillnames to Developers}
    \vspace{-2.5mm}
\end{figure}

Next, we study English-Speaking Markets (ESM) alone. 
Table \ref{tb:Cloned Skills} shows the number of skills that reuse names {by market category in ESM. }
Games and Trivia turn out to be the category with the highest number of reused skill invocation names.
We also observe 968 skills with the same skill invocation name and developer name in a single market (IN-280, CA-121, DE-36, UK-192, ES-14, FR-8, US-279 , AU-38). One good example is the ``Africa Facts'' in the UK market by \textit{Yasemin Woodward} with skill ids B07PGN7T9D, B07PHRMML4, B07PK4GNYC, B07PKVMYTB, B07PR67XTQ, and B07PFQ39Y6. This suggests some developers might be attempting a form of \emph{sybil attack}~\cite{douceur2002sybil}, where they try to have multiple skills with the same invocation name (but different ID) to increase the chance for one of them to be selected.

\begin{table}[ht]
\small{
\centering
\caption {Number of reused skill invocation names per category in the English-speaking markets.}
\label{tb:Cloned Skills}
\scalebox{.8}{
\begin{tabular}{m{2.8cm}|ll|ll}
Category & Invocation names & \% & Skills & \%  \\ \hline
Games \& Trivia & 7,251 & 30\% & 26,580 & 29\% \\
Education \& Ref. & 4,250 & 17\% & 13,863 & 15\% \\
Music \& Audio & 3,814 & 16\% & 12,121 & 13\% \\
Lifestyle & 1,697 & 7\% & 6,074 & 7\% \\
Smart Home & 1,130 & 5\% & 3,456 & 4\% \\
Health \& Fitness & 977 & 4\% & 3,505 & 4\% \\
Food \& Drink & 597 & 2\% & 2,111 & 2\% \\
Business \& Finance & 783 & 3\% & 2,494 & 3\% \\
Kids & 683 & 3\% & 2,356 & 3\% \\
Others & 3,286 & 13\% & 15,235 & 22\% \\ \hline
\textbf{Total} & 24,468 & 100\% & 92,451 & 100\% \\\hline
\end{tabular}
}}
\end{table}

Figure~\ref{Fig:Popular Skills Name} shows the most popular skill invocation names in the 5 English-speaking marketplaces. 
``Cat Facts'' is the most popular skill invocation name used 171 times. 
Likewise, the word ``Fact'' is used by {1,979 developers appearing 3,905 times across all marketplaces.}
We observe that many of the fact skills are single interaction skills (they terminate their interaction after performing one task), out of which over 40\% are developed using automated platforms. 
This may be related to these platforms offering free fact skill templates as discussed above. %

\begin{figure}[ht]
    \centering
    \includegraphics[width=.65\columnwidth,trim=17 7 17 10,clip]{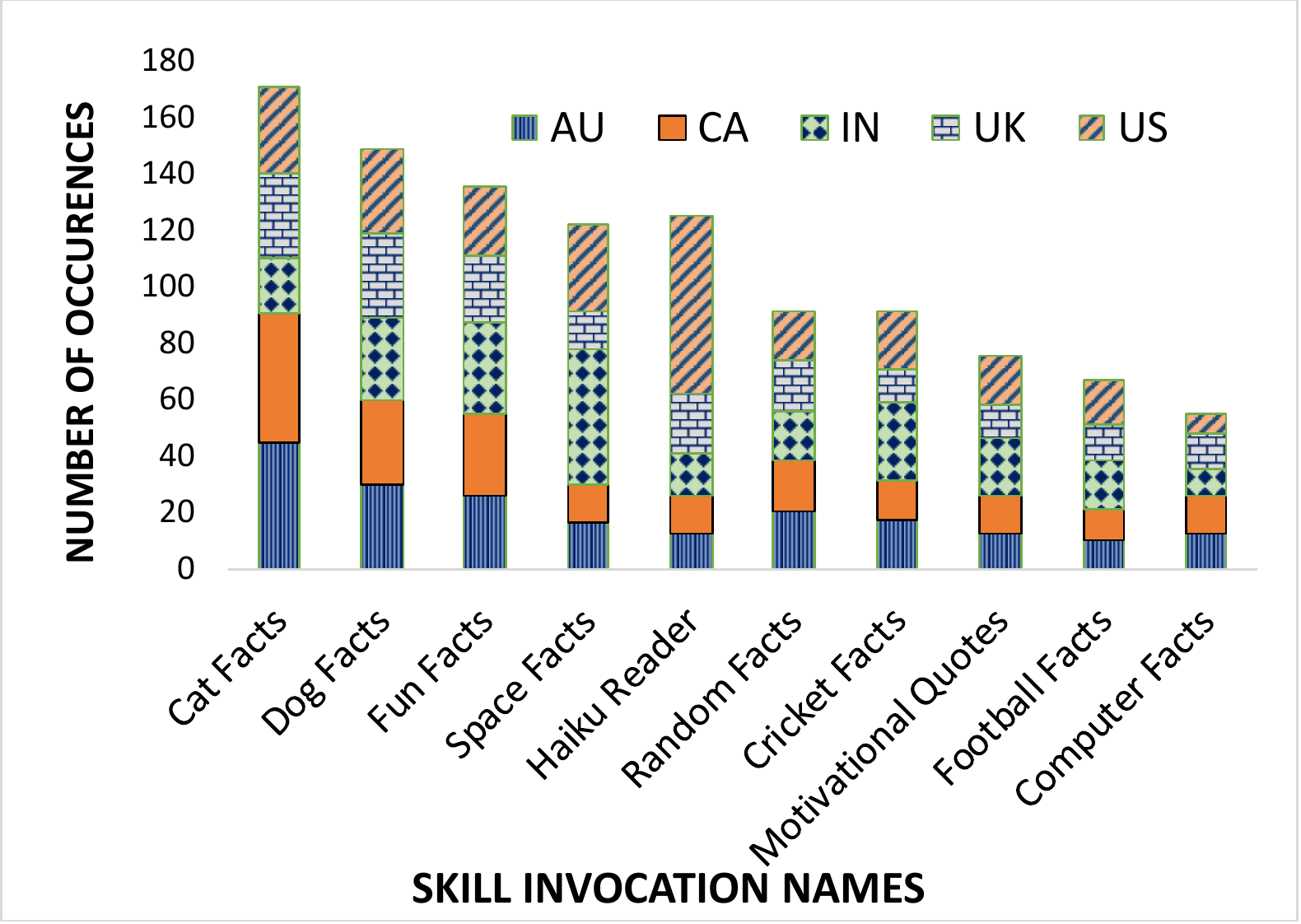}
            \vspace{-7pt}
    \caption{Top 10 skills invocation names (English-speaking).}
    \label{Fig:Popular Skills Name}
    \vspace{-2.5mm}
\end{figure}

\subsection{Invocation Name Similarity}
\label{sec:phoneticsim}

Because of the slight differences in invocation name, we also sought to measure how similar skill invocation names are. 
To do this, we use the Levenshtein edit distance~\cite{Lev65} to compute the similarity between skill invocation names. 
We also consider the phonetic similarity between invocation names~\cite{191968}, i.e., the phonetic transcription of an invocation name. This is because both textual and phonetic invocation name similarity may be leveraged to exploit the speech recognition capabilities of SPA for an impersonation attack, known as skill squatting~\cite{zhang2019dangerous, 217575}. 

We use the open-source Carnegie Mellon University Pronouncing Dictionary (CMUdict) to get the phonetic transcription of words.\footnote{\url{https://github.com/cmusphinx/cmudict} (version 0.7b).} 
For every skill in every English-speaking market, we lowercase its invocation name, remove all non-ASCII characters and punctuation, and replace digits with their equivalent texts (e.g., ``1'' becomes ``one''). Afterwards, we use the CMU dictionary to obtain the phonetic transcription of every skill invocation name, ignoring those skill invocation names without phonetic transcription.
For every resulting skill, we estimate the shortest phonetic Levenshtein distance to any other skill invocation name in the same market to identify skills with similar sounding invocation names. In the process we factor in the dynamic lengths of different skills invocation name by normalizing the Levenshtein distance by the length (in phonemes) of the longest skill invocation name compared.

Table~\ref{tb:name Levenshtein distance} shows the quantity and percentage over the total number of skill invocation names we found with a phonetic translation and a minimum edit distance when compared to the other skills of $\leq$ 0.1 and $\leq$ 0.2 across the five English-speaking marketplaces. We observe that in all English-speaking marketplaces (except India) we have an average of 6 to 8\% skills invocation names with minimum Levenshtein distance $\leq$ 0.1, and around 26\%  with minimum distance $\leq$ 0.2. Most of the skills with lower phonetic Levenshtein distances $\leq$ 0.1 are plural versions of the original skill, such as ``Fun Fact Quiz'' and ``Fun Facts Quiz'' (India), or ``Panda Facts'' and ``Panda Fact'' (US). However, there are others%
, such as ``Sweet Facts'' and ``Wheat Facts'' (US), which would have a higher edit distance without transcription but that when compared after transcription have a Levenshtein distance $\leq 0.1$. %
In either case, we can see that there is a considerable amount of skill invocation names that are quite similar to each other across the five English-speaking markets.
For instance, we see that in the US, 5\% of skills (3,215) have a Levenshtein edit distance (after phonetic transcription) $\leq$ 0.1, and 15\% of skills (7,332) have a distance $\leq$ 0.2\%. 

\begin{table}[t]
    \centering
    \small{
    \caption{skill invocation names with Levenshtein distance $\leq 0.2$ across English-speaking markets. }
    \label{tb:name Levenshtein distance}
    \scalebox{0.9}{
    \begin{tabular}{l|r|r |r}
       Market & \emph{lev} $\leq$ 0.1 & \emph{lev} $\leq$ 0.2 & Total      \\ \hline
    US          & 3,830 (8.82\%)      & 11,929 (27.47\%)                &   43,410   \\ \hline
    CA      & 1,514 (7.88\%)      & 4,836 (25.17\%)                 &   19,209   \\ \hline
    AU   & 673 (6.86\%)        & 4,439 (24.69\%)                 &   17,974   \\ \hline
    IN       & 3,048 (14.25\%)     & 6,699 (31.31\%)                 &   21,389   \\ \hline
    UK          & 1,714 (7.80\%)      & 5,321 (24.24\%)                 &   21,950   \\
    \end{tabular}}
    }
    \vspace{-3.0mm}
\end{table}
}

\ifthenelse{\boolean{longver}}{\subsection{Characterization Take-aways}}{\noindent\textbf{Take-aways.}} Our market characterization provides a high-level overview of the skill ecosystem and how Alexa marketplaces are structured. 
In summary, the main takeaways %
include:
\begin{compactitem}%
    \item The skill ecosystem has grown considerably fast in recent years and it currently has few developers with thousand of skills each. 
    English-speaking markets generally dominate and push skills to other markets. 
    
    \item We see skills developed with automated platforms. 
    The use of these platforms lower down the development barriers and enable bulk skill creation. 
    
     \item We observe a high prevalence of skills with the same or  similar name, with the associated potential for impersonation and Sybil attacks as shown in previous works.
    
\end{compactitem}

\section{Permissions and Traceability}
\label{sec:permissions}

In this section, we first explore the data permissions in all Alexa marketplaces, analyzing which permissions are more often requested by Alexa skills and developers among the different markets, and how they are distributed between skills. 
Then, we look at skill privacy policies to understand how developers disclose and justify the data permissions they ask for.
To do so, we {create} the \textbf{largest traceability dataset for Alexa} known to date, namely the traceability-by-policy dataset (TBPD), which includes the traceability analysis of 1,758 skills requesting data permissions and their policies from the five English-speaking marketplaces (Australia, UK, US, Canada, and India), representing 67.4\% of all skills requesting data. 
Using the dataset, we analyze the skill traceability at the skill category and developer dimensions and identify developers' %
practices.

\subsection{Data Permissions}
\label{subsec:data-permissions}
Skills can request access to user information through the Alexa skill API. 
This information becomes available on a per-skill and per-data-type basis when the user consents.
To manage the consent: i) Alexa declares the permissions of skill on their store and gives users control through the Web or the mobile app, %
and ii) skills have to inform users in their privacy policies how they will use the personal information they collect. 
Alexa currently supports 11 types of permissions~\cite{Alexa_permissions_list}. These are: 
Device Address (15.0\%), 
Location Services (4.1\%), 
Email Address (15.7\%), 
Device Country and Postal Code (13.7\%), 
Reminder (9.1\%), 
Customer Name (10.3\%), 
List Read Access (5.3\%), 
List write Access (4.9\%), 
Mobile Number (4.3\%), 
Amazon Pay (1.8\%), 
and Skill Personalization (0.0\%).\footnote{At the time of our analysis, we do not see any skills with Skill Personalization yet. Also, our analysis reveals Notification (15.5\%), which is now deprecated and replaced by Reminder and Timers (0.3\%).} %
The most prevalent permissions are generally used to offer services based on the user's location. 
\ifthenelse{\boolean{longver}}{Figure~\ref{Fig:Data Permissions distribution across the marketplaces} lists the 13 Alexa permissions together with the number of skills that use them per market.}

\ifthenelse{\boolean{longver}}{
\begin{figure}[ht]
    \centering
        \vspace{-5pt}
    \includegraphics[width=1\columnwidth, trim=11 30 22 10, clip]{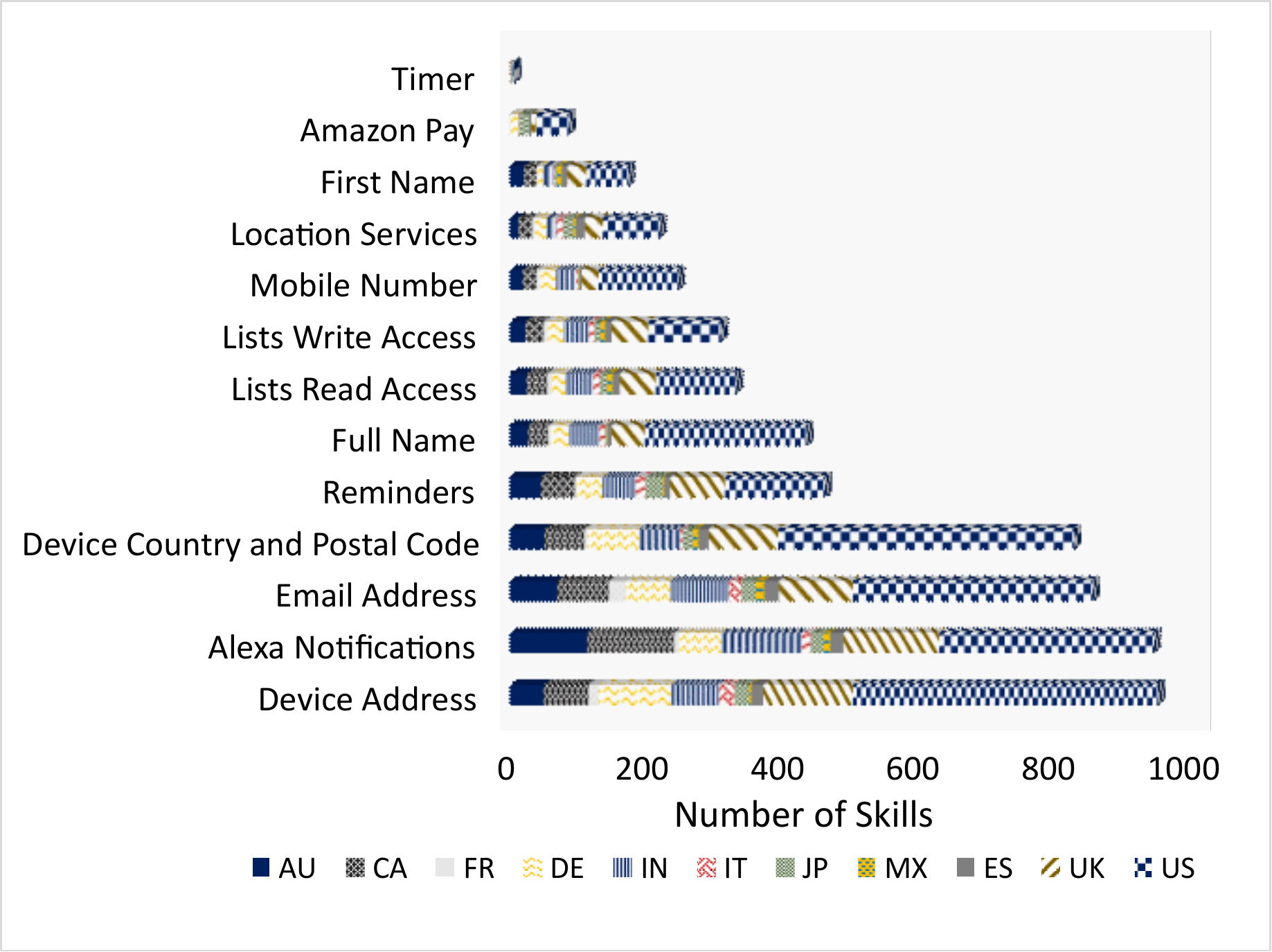}
    \vspace{-17pt}
    \caption{Permissions distribution across the markets.}
    \label{Fig:Data Permissions distribution across the marketplaces}
    \vspace{-2.0mm}
\end{figure}
}

Overall, 2,608 (2.3\%) skills across markets ask for data access permissions. This could, in principle, be seen as good news. Still, 2.6K skills is a sizable set of third-party software accessing personal information, and the practices of their developers require scrutiny.  
We also argue that skills may collect personal information using other means, i.e., through account linking or conversations, as discussed in \S\ref{sec:bypassing}. %

\begin{table}[t]
	\centering
	\caption {Permissions distribution by skills and marketplace.}
	\label{tb:Total unique Permissions ask by different skills}
	\scalebox{0.8}{
		\begin{tabular}{|c|c|c|c|c|c|}
			\multirow{3}{*}{\textbf{Skills}} & \multicolumn{5}{c}{\textbf{No of Unique Permissions}} \\
			&  1              & 2           & 3          & 4         & $\geq$5            \\
			& N (\%)          & N (\%)      & N (\%)     & N (\%)    & N (\%)             \\
			\hline
			US (1,698) & 1,169 (68.85) & 301 (17.73) & 143 (8.42) & 51 (3.00) & {34 (2.00)} \\
			UK (581) & 418 (71.94) & 120 (20.65) & 26  (4.48) & 7  (1.20) & {10 (1.72)} \\
			CA (366) & 262 (71.58) & 74  (20.22) & 14 (3.83) & 6  (1.64) & {10 (2.7.73)} \\
			IN (358) & 250 (69.83) & 66 (18.44) & 20  (5.59) & 7  (1.96) & {15 (4.19)} \\
			AU (348) & 245 (70.40) & 78  (22.41) & 10  (2.87) & 5  (1.44) & {10  (2.87)} \\
			DE (341) & 236 (69.21) & 83  (24.34) & 12  (3.52) & 7  (2.05) & {3  (0.88)} \\
			JP (126) & 100 (79.37) & 21  (16.67) & 4  (3.17) & 1  (0.79) & {} \\
			ES (89) & 56  (62.92) & 24  (26.97) & 6   (6.74) & 3  (3.37) & {} \\
			IT (85) & 58  (68.24) & 21  (24.71) & 5   (5.88) & 1  (1.18) & {} \\
			FR (63) & 44  (69.84) & 15  (23.81) & 4   (6.35) &  & {} \\
			MX (46) & 30  (65.22) & 13  (28.26) & 3   (6.52) &  & {} \\ \hline  
			\textbf{Total} 4,101 & 2,868 (69.93) & 816 (19.90) & 247 (6.02) & 88 (2.15) & {82 (2.00)} \\
			\textbf{Uniq.} 2,608 & 1,807 (69.29) & 499 (19.13) & 191 (7.32) & 67 (2.57) & 44 (1.69) \\
		\end{tabular}
	}
\end{table}

Table \ref{tb:Total unique Permissions ask by different skills} shows the permission distribution of skills per marketplace. As shown, most skills ask for one permission (typically, the device address as discussed above), with all marketplaces showing a similar pattern overall --- albeit with slight differences between those markets with more or less number of skills. 
There is also an important minority of skills asking for {two or} three permissions.
Finally, there are few exceptions where skills ask for over three permissions. 
The most noticeable one is a skill called ``BILH Staff Chatbot'' by \emph{BIDMC} (the Beth Leahy hospital). 
The skill is available in the US, and it is requesting several unique permissions (i.e., Name, Email Address, Mobile Number, Reminders, Location Services, and Device Country and Postal Code). 
The skill is a chatbot for employees to help them fill out a COVID-19 staff daily health questionnaire. While the skill is not intended to diagnose users (according to the description), it asks for 13 different symptoms. 
The use of this skill poses an important privacy risk to employees,
because the Web form version of the questionnaire does not ask for data such as postal code or location, so it is not clear why these are needed in Amazon.

\subsection {Traceability Analysis}\label{sec:pt}
\label{subsec:traceability}
We look at privacy policies to understand how developers disclose the permissions they request. Specifically, we study the traceability between data operations obvious to users (recall users need to enable skills requesting permissions)  
and the data actions defined by developers in the skill policies. 
Note that Amazon's privacy requirements for skill developers mandate that a skill must come with an adequate privacy policy if it collects personal information. In particular, any collection and use of personal information need to comply with what is stated in the privacy policy \cite{alexa_security_requirement}.

To evaluate traceability between permissions and policies, we collect and tag privacy policies of 1,758 Alexa skills requesting data permissions in the five English-speaking marketplaces.
We develop a Selenium module in Python that automatically visits and downloads every skill's privacy policy page, cleaning the HTML code to remove unnecessary markup code, normalizing punctuation, and extra white-spaces discarding non-ASCII characters, and finally converting text to lowercase (this is later refered as the pre-processing phase).  
Afterwards, each skill is analyzed and evaluated as having \emph{broken}, \emph{partial} or \emph{complete} traceability, following previous studies of traceability analysis in other domains \cite{young2010method,anthonysamy2013social,misra2017privacy,Sebastian_Sebastian_Zimmeck_Peter}.

The type of traceability is identified by comparing the permissions requested by the skill through the Amazon Alexa API with the data practices covered in the skill policy. 
The different traceability types are explained next:

\noindent\textit{\uline{Complete:}}
A skill is said to offer complete traceability if it provides adequate information in its privacy policy document about its data practices, i.e., the data action defined in the privacy policy document can be completely mapped to the access of data permissions. 
{For instance, a skill like the ``Aircraft Radar'' in the UK market developed by \emph{Chris Dzombak} offers complete traceability since it provides adequate information about its data practices in its privacy policy. 
The statement ``aircraft radar uses your devices address to find your location and search for aircraft around you'' can be mapped with the Device Address permission the skill collects}. %

\noindent\textit{\uline{Partial:}}
This is when the transparency of data action defined in the privacy policy document maps partially to the access permissions. For example, when the privacy document states that a skill collects personally identifiable information without explicitly stating what this information is. 
A statement such as ``we may require you to provide us with certain personally identifiable information'' offers partial traceability since it does not explicitly state what information is collected. 
A skill is also said to offer partial traceability if not all its data permissions are covered in its privacy document. Partial disclosure of data practices may also occur when data practices in a skill privacy document are not well mapped with the skill's data permission. For instance, the statement ``we may also collect your zip code'' is partially traceable as the skills collect Device Address, and this refers to the user's full address, including the zip code and the street number. 
{Examples that provide partial information in their privacy policy include: 1) ``Vote Sam Feldt'' by \emph{Fanspoke} that collects Mobile Number 2) ``Flight Level - An Aircraft Radar'' by \emph{O. Schafer} that collects Device Country and Postal Code.}  

\noindent\textit{\uline{Broken:}}
If a skill requests for permissions but does not have a privacy policy or the link to the privacy policy given is not working, we consider the traceability broken. Even when providing a privacy policy, A skill is said to have broken traceability if it provides no data implication in its privacy policy document. For instance, when a skill collects the device address, and its privacy document states nothing related to device address, we mark such skill data practices disclosure as broken.
We only mark a broken policy when permissions are not traceable to the data practices defined in the privacy document. 
In the UK marketplace, skills like ``Daily Yoga'' by \emph{Siva Pandeti} collects the Device Country and Postal Code, ``Bike Weather Man'' by \emph{Marc Easen} collect Device Country \& Postal Code, and  ``Smoggy Alerts'' by \emph{Teal Dreams Software, Inc.} collects the Device Country and Postal Code, and they all exhibit broken traceability.

When analyzing traceability, a skill requesting ``Device Country and Postal Code'' permission is tagged to have adequately disclosed its data practices if ``Device Address'' is mentioned in the privacy policy. This is because ``Device Address'' refers to the user's full address, including the device country and postal code and the street number. Also, we did not find any skills using Skill Personalization, and grouped %
``List Write Access'' and ``List Read Access'' under the ``Personal Information'' category as acknowledging personal information collection should be sufficient to disclose these permissions.
Finally, reminders/notifications are not explicitly considered in the privacy requirements for skills by Amazon, as they may not have a personally identifying implication.
Therefore, the traceability of all skills was evaluated considering: ``Location Services'', ``Amazon Pay'', ``Mobile Number'', ``Email Address'', ``Name'', ``Device Address'', ``Device Country and Postal Code'', ``Personal Information''.

\subsection{Traceability Results} \label{subsec:traceability_results}

\begin{figure}[ht]
    \centering
    \includegraphics[width=.7\columnwidth,trim=2 5 10 8,clip]{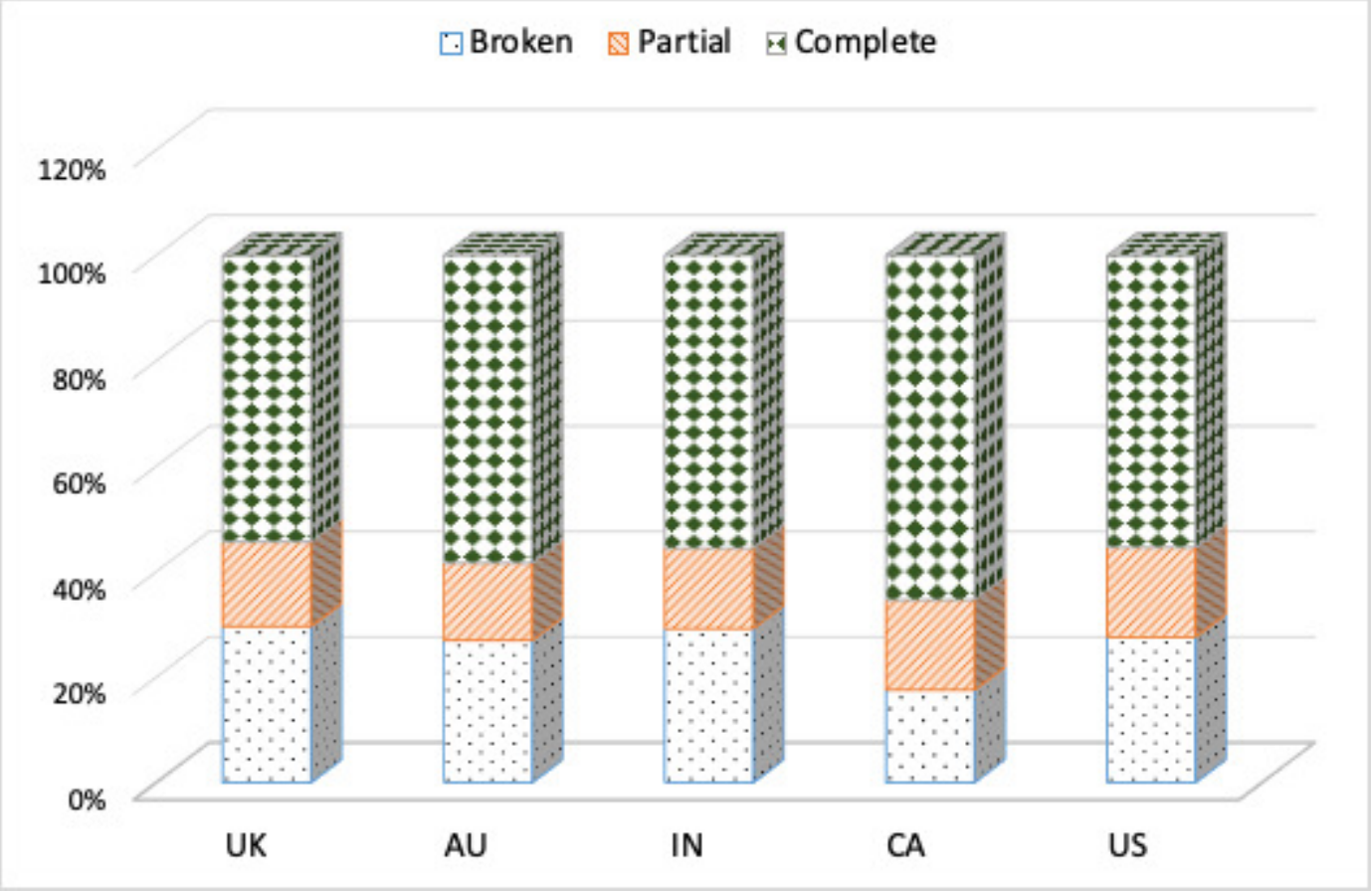}
    \vspace{-4.0mm}
    \caption{Traceability for skills requesting data permissions in English-speaking markets.} 
    \label{Fig:Traceability-Results-English-Marketplaces}
\end{figure}

A total of 1,758 skills request data permissions in English-speaking marketplaces. Out of the 1,758 skills: 
442 (25\%) have broken traceability, 
306 (18\%) have partial traceability, and
1,010 (57\%) have complete traceability.
When we exclude the complete ones, we see 43\% of the skills displaying bad privacy practices. 
Figure~\ref{Fig:Traceability-Results-English-Marketplaces} shows the results by market, where we can see a very similar trend across marketplaces, with UK appearing to have slightly more broken and less complete traceability results, but it also has the least number of skills and developers.

\noindent\textbf{Traceability by Categories.} 
To understand how traceability affects different type of skills, we first compare our results considering the 1,758 unique skills collected across different skill subcategories and English-speaking marketplaces, as shown in Table~\ref{tb:traceability by categories}. We show the breakdown using subcategories to give a more fine-grained perspective of particular types of (or sector-specific) skills and associated traceability. 
We rank the different subcategories based on the number of issues (broken and partial) normalized by the number of well defined policies (complete). Note that we do not list subcategories with less than 3 skills for the sake of clarity.

\begin{table}[ht]
\footnotesize
\renewcommand{\arraystretch}{0.9}
\centering
\caption{Traceability per subcategory %
in English-speaking markets.}
    \label{tb:traceability by categories}
            \vspace{-4pt}
 \scalebox{0.8}{
\begin{tabular}{m{3cm}|c|c|c|c}
\textbf{Categories} &  \textbf{R} & \textbf{B} & \textbf{P} & \textbf{C} \\ 
\hline
Sports & 1 & 43 (91.49\%) & 1 (2.13\%) & 3 (6.38\%) \\ 
Social & 2 & 98 (85.96\%) & 7 (6.14\%) & 9 (7.89\%) \\ 
(Uncategorised) & 3 & 56 (90.32\%) & 1 (1.61\%) & 5 (8.06\%) \\ 
Novelty \& Humour & 4 & 5 (71.43\%) & 2 (28.57\%) & 0 (0.00\%) \\ 
Kids & 5 & 10 (83.33\%) & 1 (8.33\%) & 1 (8.33\%) \\ 
Games \& Trivia & 6 & 89 (57.42\%) & 34 (21.94\%) & 32 (20.65\%) \\ 
Schools & 7 & 3 (100.00\%) & 0 (0.00\%) & 0 (0.00\%) \\ 
Utilities & 8 & 17 (56.67\%) & 5 (16.67\%) & 8 (26.67\%) \\ 
News & 9 & 30 (63.83\%) & 4 (8.51\%) & 13 (27.66\%) \\ 
Health \& Fitness & 10 & 60 (48.78\%) & 25 (20.33\%) & 38 (30.89\%) \\ 
Organizers \& Assistants & 11 & 7 (63.64\%) & 1 (9.09\%) & 3 (27.27\%) \\ 
Lifestyle & 12 & 74 (46.54\%) & 32 (20.13\%) & 53 (33.33\%) \\ 
Weather & 13 & 22 (41.51\%) & 13 (24.53\%) & 18 (33.96\%) \\ 
Food \& Drink & 14 & 39 (42.86\%) & 20 (21.98\%) & 32 (35.16\%) \\ 
Streaming Services & 15 & 4 (57.14\%) & 1 (14.29\%) & 2 (28.57\%) \\ 
Productivity & 16 & 42 (51.85\%) & 9 (11.11\%) & 30 (37.04\%) \\ 
Business \& Finance & 17 & 82 (52.23\%) & 16 (10.19\%) & 59 (37.58\%) \\ 
Smart Home & 18 & 44 (57.14\%) & 4 (5.19\%) & 29 (37.66\%) \\ 
Travel \& Transportation & 19 & 28 (37.33\%) & 18 (24.00\%) & 29 (38.67\%) \\ 
Education \& Reference & 20 & 59 (43.70\%) & 16 (11.85\%) & 60 (44.44\%) \\ 
Movies \& TV & 21 & 5 (50.00\%) & 1 (10.00\%) & 4 (40.00\%) \\ 
Navigation \& Trip & 22 & 2 (40.00\%) & 1 (20.00\%) & 2 (40.00\%) \\ 
Connected Car & 23 & 9 (39.13\%) & 3 (13.04\%) & 11 (47.83\%) \\ 
Public Transportation & 24 & 2 (16.67\%) & 4 (33.33\%) & 6 (50.00\%) \\ 
Novelty \& Humor & 25 & 3 (25.00\%) & 3 (25.00\%) & 6 (50.00\%) \\ 
Home Services & 26 & 5 (19.23\%) & 6 (23.08\%) & 15 (57.69\%) \\ 
Self Improvement & 27 & 2 (50.00\%) & 0 (0.00\%) & 2 (50.00\%) \\ 
Wine \& Beverages & 28 & 1 (20.00\%) & 1 (20.00\%) & 3 (60.00\%) \\ 
Music \& Audio & 29 & 94 (29.38\%) & 5 (1.56\%) & 221 (69.06\%) \\ 
Shopping & 30 & 26 (20.63\%) & 10 (7.94\%) & 90 (71.43\%) \\ 
Calendars \& Reminders & 31 & 1 (14.29\%) & 1 (14.29\%) & 5 (71.43\%) \\ 
Pets \& Animals & 32 & 1 (33.33\%) & 0 (0.00\%) & 2 (66.67\%) \\ 
Event Finders & 33 & 1 (25.00\%) & 0 (0.00\%) & 3 (75.00\%) \\ 
Local & 34 & 0 (0.00\%) & 1 (25.00\%) & 3 (75.00\%) \\ 
Knowledge \& Trivia & 35 & 1 (8.33\%) & 0 (0.00\%) & 11 (91.67\%) \\ 
\hline
\end{tabular}
}
\footnotesize
\newline
\textbf{B} = Broken, \textbf{P} = Partial, \textbf{C} = Complete, 
\textbf{R} (Rank) $\sim$ (B+P)/(C+1)
\end{table}

Sports has the largest proportion of skills with broken traceability, totaling 91\% of all the skills in the subcategory. %
For instance, the ``Western States'' skill by \emph{de Peck, Inc} collects the Device Address without any privacy document to disclose the data practices.
Like many other subcategories, there are no skills with complete traceability in Novelty \& Humour. 
When looking at the combination of broken and partially broken traceability, Games \& Trivia is also one of the most problematic subcategories, particularly considering the number of skills it has.
For instance, we find ``Pixated Salat'', a prayer skill ``to find Salah or Prayer time''. 
The skill is developed by \emph{Pixated ltd}, a strategic advertising group, and asks for the device's location. However, the link to the policy provided is broken, and it links to the main site of the advertising group instead (\url{http://pixated.agency}).

It is also important to note that even in categories that do not rank high, there are also skills with concerning data collection practices, like the Education \& Reference category. This subcategory has a high proportion of complete traceability (44\%), but it also has a very sizable proportion of broken traceability skills (43\%). For instance, in this category, we find a very interesting case of ``A Sales Guy'' by \emph{VoiceXP}. 
This skill is supposed to provide information about Keenan, that according to his website, is a person who has been ``selling something to someone for his entire life''. 
Users that want to know about Keenan would have to give up their mobile number, email address, full name, or device address. 
The developer, \emph{VoiceXP}, is a company that does so-called {\it voice domain registration services}. 
We refer the reader to \S\ref{sec:DISCUSSION} for further discussions on the type of concerning practices used by skill developers.

Regarding complete traceability, it is worth highlighting the bottom part of Table~\ref{tb:traceability by categories}. Particularly, the \emph{Music \& Audio} subcategory has the largest number of complete traceability skills (221), which is also a sizable proportion of skills within the subcategory (69\%). This is closely followed by Shopping, with many complete traceability skills (90) that make up 71\% of skills within that subcategory. 
Note that these two categories relate to industries with a larger tradition of offering services in the Web, where privacy has been in scrutiny for longer. 
Still, adding up the skills with broken and partial traceability across the two subcategories gives a total of 135 skills. Therefore, even the best ranking categories have a sizable number of broken/partial traceability skills.

Finally, and rather interestingly, the Schools and Kids subcategories are ranked in positions \#7 and \#5, and have 100\% and 0\%, and 83\% and 8\% of their skills broken or partially broken, respectively, for a total of 11 skills. This is especially concerning as they may collect children's personal information, as we discuss more thoroughly in \S\ref{sec:DISCUSSION}.

\noindent\textbf{Traceability by Permissions.}
To understand how traceability varies across the different type of permissions, we also look at the traceability of skills per permission requested.
Table \ref{tb:traceabilitybypermissionall} shows the distribution of traceability across different permissions for the 1,758 analyzed skills. 
The permissions are first grouped into  Broken, Partial, Complete, with respect to the policies of the skills where these permissions are requested. A total of 2,616 permissions are requested (622 by skills with broken traceability, 485 by skills with partial traceability (75 of these are traceable), and 1,509 by skills that exhibit complete traceability). The most requested permission is the \emph{Device Address} which is requested 648 times by 464 developers. While \emph{Amazon Pay} is the least asked permission requested only 56 times by 33 developers, it tends to be requested more by skills that have complete traceability. %
In contrast, \emph{Location Services} permission requested by 90 unique developers is found more in skills that exhibit broken traceability.

\begin{table}
\centering
\caption{Distribution of traceability across different permissions for the 1,758 analyzed skills in the 5 English-speaking marketplaces.}
\label{tb:traceabilitybypermissionall}
\scalebox{0.86}{
\begin{tabular}{l|l|c|l|l|l}
Permission                  & D     & R     & B         & P          & C           \\ 
\hline
Device Address              & 464   & 648   & 188 (29\% )& 130 (20\% ) & 330 (51\%)   \\
Device Country              & 330   & 569   & 106 (19\% )& 77 (14\%)   & 386 (68\% )  \\
Email Address               & 251   & 428   & 99 (23\% ) & 77 (18\%)   & 252 (59\% )  \\
Personal Info.              & 144   & 324   & 67 (21\% ) & 41 (13\%)   & 216 (67\% )  \\
Name                        & 173   & 350   & 86 (25\% ) & 82 (23\%)   & 182 (52\% )  \\
Mobile Number               & 97    & 139   & 36 (26\% ) & 37 (27\%)   & 66 (47\% )   \\
Location Services           & 90    & 102   & 35 (34\% ) & 31 (30\%)   & 36 (35\% )   \\
Amazon Pay                  & 33    & 56    & 5 (9\% )   & 10 (18\%)   & 41 (73\% )   \\ 
\hline
\multicolumn{1}{r|}{Total}  & 1,582 & 2,616 & 622 (24\% )& 485  (19\% )& 1,509 (58\% ) \\
\multicolumn{1}{r|}{Unique} & 1,123 & 1,758 & 442 (25\% )& 306  (18\% )& 1,010 (57\% ) \\\hline
\end{tabular} 
}
\\ \textbf{D} = Developer, \textbf{R} = Requested, \textbf{B} = Broken, \textbf{P} = Partial, \textbf{C} = Complete, \\ %
\end{table}

\noindent\textbf{The Good, the Bad and the Ugly Developers.} 
We next study how many ``good'', ``bad'' and ``average'' developers there are. 
Table~\ref{tb:developers_traceability} shows the number of developers per type of traceability considering the 5 English marketplaces. 
Overall, we see a total aggregate across markets of 1,730 cases where developers request permissions in their skills, out of which 1,123 are unique developers that post skills in several markets. 
When looking at unique developers, we see:

\noindent{\em \uline{The Good}}.
There are 566 developers with all their skills showing complete traceability. All the skills developed by these developers have statements in their privacy policies clearly stating and justifying the permissions they request. This accounts for about {50\%} of the developers.
For instance, developers such as \emph{GoVocal.AI}, \emph{Blutag Inc}, and \emph{Ixartz} in the US marketplace have complete traceability between the permissions their skills ask and their privacy policies. 

\noindent{\em \uline{The Bad}}.
There are 350 developers with all their skills broken. This accounts for about {31\%} of the developers.
Skills do not offer an adequate explanation in general when we analyze the partial skills, their privacy statements, and we look at their reviews. 
One example is \emph{Tagrem Corp} in the US market, with all the skills they developed exhibiting broken traceability, as it does not acknowledge the collection of any personal information in the skills' privacy policy while the skills do request for permissions.

\begin{table}
\centering
\caption{Developers' disclosure practices (Broken, Partial or Complete) in all 5 English-speaking marketplaces.}
\label{tb:developers_traceability}
\begin{tabular}{c|c|c|c|c|l} 
Markets & Developers & B & P & C & P+C \\ 
\hline
US & 731 & 219 & 145 & 330 & 15 \\
UK & 402 & 197 & 41 & 146 & 7 \\ 
CA & 224 & 88 & 27 & 102 & 2 \\ 
IN & 215 & 107 & 23 & 79 & 5 \\ 
AU & 202 & 90 & 22 & 82 & 3 \\ 
\hline
\multicolumn{1}{r|}{Total} & 1,730 & 701 & 258 & 739 & 32 \\ 
\multicolumn{1}{r|}{Unique} & 1,123 & 350 & 207 & 566 & 21 \\
\hline
\end{tabular}
    \\ \textbf{D} = Developer, \textbf{B} = Broken, \textbf{P} = Partial, \textbf{C} = Complete
    \vspace{-2.0mm}
    \end{table}

\noindent{\em \uline{The Ugly}}.
We see 207 developers with all their skills with partial traceability. 
This accounts for about {18\%} of the developers. They appear to have a sloppy attitude when writing privacy policies and informing users of how the personal information they request is used.
For example, the developer \textit{Geekycoders} with over 30 skills in the US market have partial traceability in all of them. While the developer requests for only one type of permission (First Name), the statement ``we do not collect your personal information when you use any part of our products, unless the app specifically lists in any pre-download description'' only offers partial traceability since it does not explicitly state what information is collected.

\vspace{1mm}
\noindent
\fbox{\parbox{0.97\columnwidth}{
This shows that both average and bad developing practices are commonplace and widespread in Alexa across marketplaces, particularly in the US. Interventions for developers, instead of just for individual skills, may also be an effective way of vetting marketplaces.}}
\vspace{1mm}

\begin{figure}[t]
    \centering
    \includegraphics[width=.8\columnwidth,trim=5 5 5 5,clip]{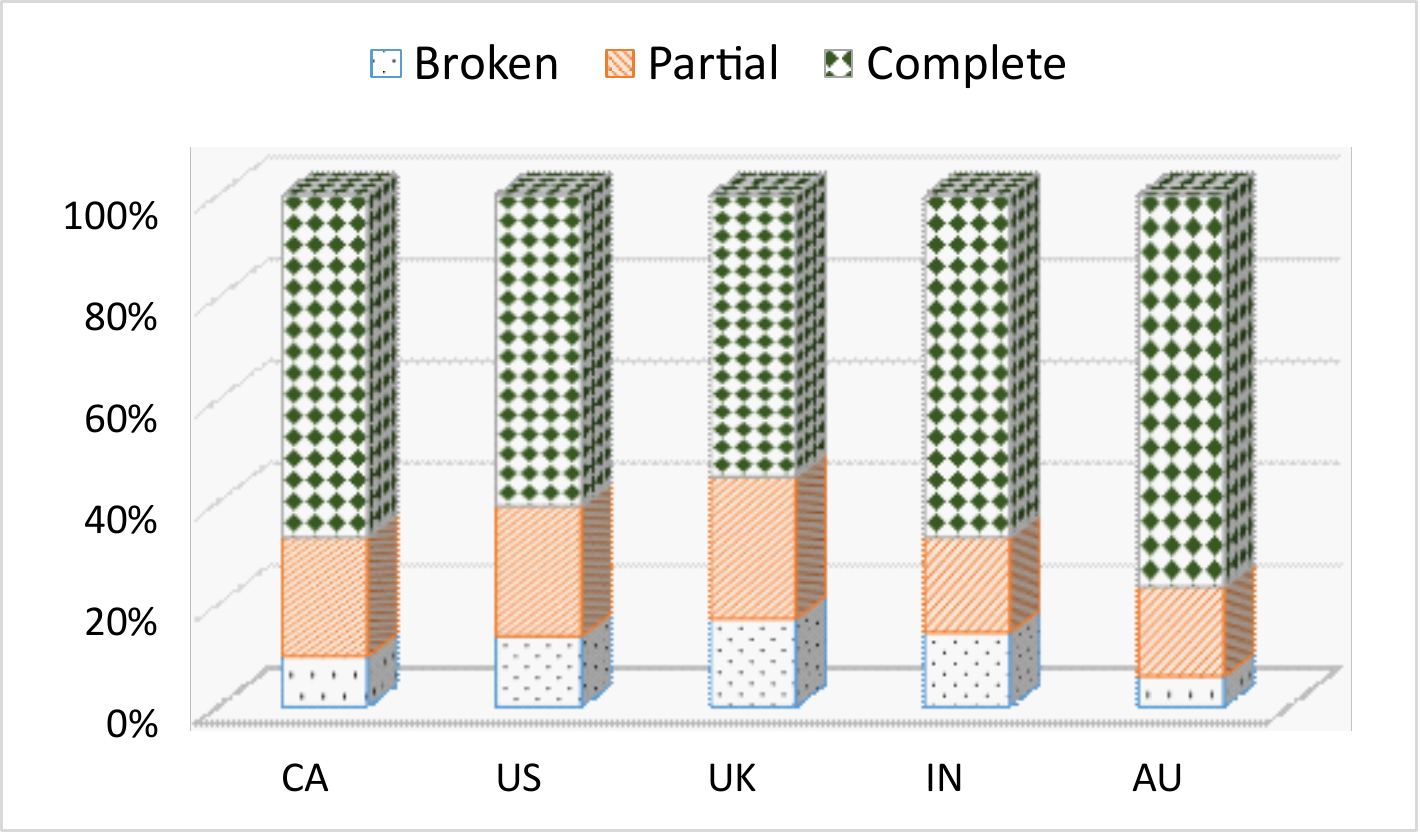}
    \vspace{-4.0mm}
    \caption{Traceability in skills reusing privacy policies (English-speaking markets). Total: 1498 skills, broken: 196, partial: 358, complete: 944. 
    }  
    \label{fg:traceability-reuse}
    \vspace{-5.0mm}
\end{figure}

\noindent\textbf{Reused Policies.} We observe that some skills are reusing the privacy policies of others.
Thus, we measure the prevalence of reused policies and study the impact this practice has on traceability by systematically mapping policy links to skills in the same English-speaking marketplace.
Figure~\ref{fg:traceability-reuse} shows an overview of our results, where 175
privacy policy links are reused across marketplaces by 1,498 skills (with the following breakdown\footnote{The breakdown is read as follows: CA-172 means 172 skills using one of the policies in the Canadian market.}: CA-172, US-825, UK-180, IN-172 and AU-149) 
from 235 developers (CA-37, AU-26, IN-9, UK-39, US-124).
Out of all the skills where we see reusing policies, only 29\% offer a comprehensive policy that adequately describes their practices, 44\% show broken traceability between policy documentation and data permissions, while 27\% show partial traceability. 
For example, the skills ``Curious City'' and ``Hits 94'' developed by \emph{Voxmatic.io} both with data permission Device Address have a privacy policy hosted in a domain that is down (\url{http://voxmatic.io/privacy}), though both are fully functional when spoken to.

Interestingly, although many of the skills with the same policy are published by the same developer and collect the same permissions, there are cases where the skill collects a different set of permissions (CA-111, AU-92, IN-109, UK-63, US-548).  
For example, in the Canada marketplace, the skills ``Big Sky'' and ``I'm Driving'' (both from developer \emph{Philosophical Creations} and with partial traceabilities) use the privacy policy link \url{https://driving.big-sky-alexa.com/privacy}. However, the first asks from Alexa Notifications and Device Address permissions while the second asks for Email Address and Mobile Number.
There are also some cases where completely different developers still reused the same privacy policy. We see 24 unique instances involving 54 different developers across all of the  English speaking markets.
For instance, in the US market, 4 developers, \emph{Evezilla Ltd, Kavson Ltd, Rai Integration Ltd, and Mikk London}, use the same policy link \url{https://www.starfishmint.com/policy/privacy.html}.

Finally, another interesting observation is that while some skills reuse broken policies and, therefore, automatically inherit the broken traceability, other skills reusing policies have different traceability due to the different permissions the skill request.
For instance, the ``Mastering Python Networking Facts'' skill by \emph{Network Automation Nerds LLC}, which asks for Device Country and Postal Code, exhibits complete traceability with the privacy policy at ``\url{https://www.alexa.com/help/privacy}''. 
However, the ``Stock Price'' skill by \emph{JJ} exhibits partial traceability when the same policy is used while asking for the Lists Read Access permission, which differs from the other skill's permissions.

\section{Automated Traceability Analysis}
\label{sec:system}

\begin{figure}[t]
\vspace{0.1cm}
    \centering
    \includegraphics[width=1.0\columnwidth, trim=40 150 190 150, clip]{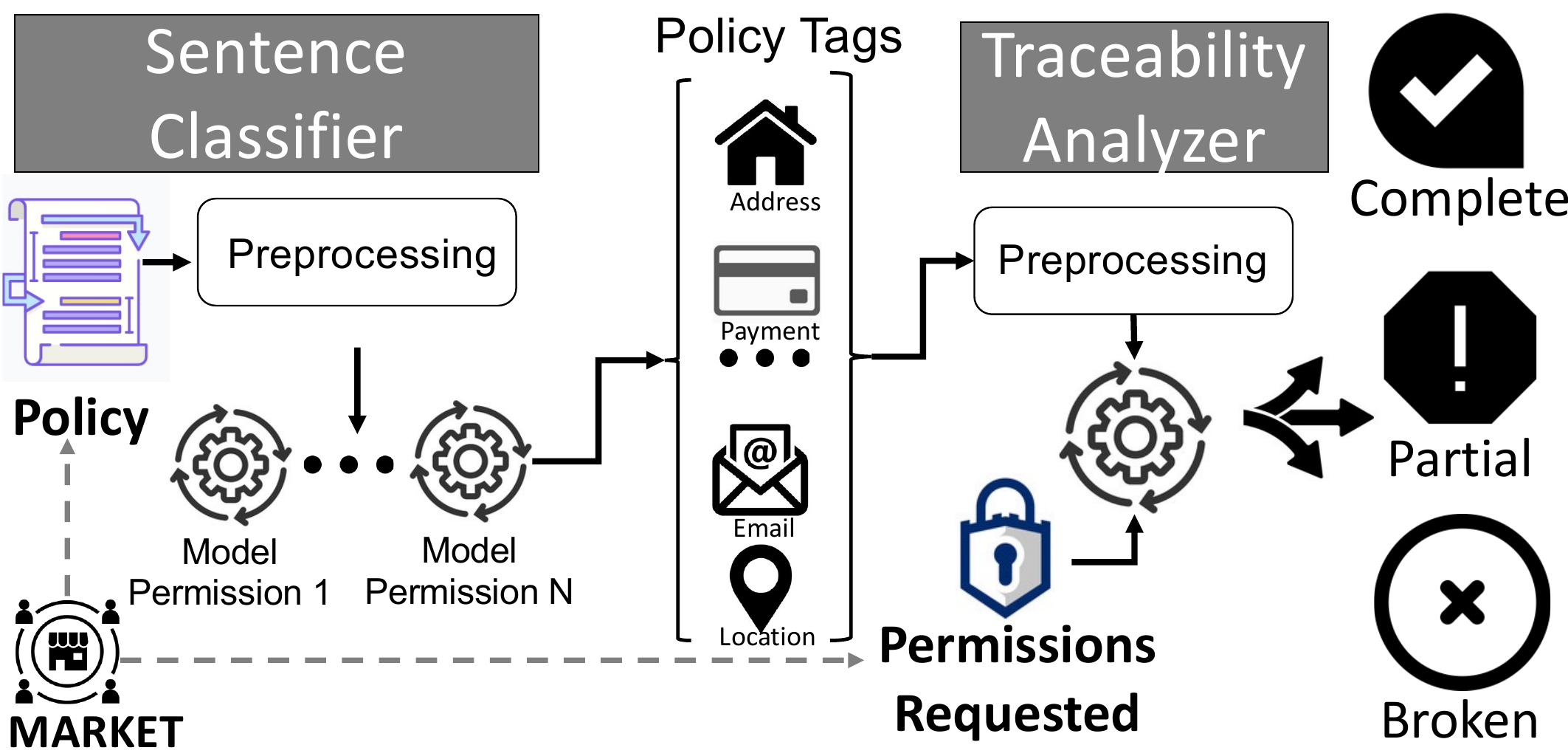}
    \vspace{-0.7cm}
    \caption{Overview of \system{}.}
    \label{fig:system}
    \vspace{-0.4cm}
\end{figure}

To facilitate the traceability analysis of skills, we create a system, \system{}, based on machine learning and natural language processing. 
Given a skill, it first identifies and classifies all statements in its privacy policy that relate to data practices over personal information. 
We map each data statement with one or more Alexa permissions. 
These are the permissions the skill justifies in the privacy policy. 
We then compare these permissions with the ones the skill is authorized to request through the Amazon API during runtime. 
Depending on if the permissions requested match those found in the policy, the skill is then classified as having a \emph{complete}, \emph{partial}, or \emph{broken} privacy policy. Figure \ref{fig:system} presents an overview of the system.
The automated traceability analysis system consists of two parts, the \emph{Sentence Classifier} and the \emph{Traceability Analyzer}. 
The Sentence Classifier (\S\ref{sec:sc}) spawns several models that perform the mapping mentioned above.  
The Traceability Analyzer (\S\ref{sec:ta}) collects all data permissions found in the privacy policy and compares them with the original set of data permissions requested by the skill. 
This is done to determine the traceability between the use of a permissions and its justification in the policy.

\subsection{Sentence Classifier}\label{sec:sc}
To identify the privacy permissions requested in each sentence automatically, we manually created the \emph{permission-by-sentence dataset (PBSD)}. 
The PBSD compiles 10,409 annotated sentences from 532 original Alexa policies randomly chosen from the TBPD described in \S\ref{sec:permissions} (the rest of TBPD is used as \emph{unseen} data for the evaluation as detailed in \S\ref{sec:syseval}). 
The annotations are tags that associate each statement to one (or more) of the Alexa permission categories of interest.  
That is, PBSD contains sentences with permissions needed for the traceability analysis as discussed in detail in \S\ref{subsec:traceability} (\emph{Amazon Pay, Device Address, Device country and postal code, Email Address, Location Services, Mobile Number, Name, and Personal Information})
and the negative label, named \emph{None}, which includes sentences that do not belong to any of the Alexa permission classes considered. 
The distribution of tagged sentences per permission category in the PBSD is presented in Table~\ref{tb:sentencecount}. 
Note the uneven distribution of sentences among classes. 
The permissions with the largest number of statements are ``Personal Information'' and ``Name''. 
Although we strive to have representative statements for all categories in Alexa, some less actionable permissions such as ``Device Address'' or other recently incorporated permissions like ``Amazon Pay'' are still not commonly used among skills, and they are therefore not easy to find and tag. 
To mitigate this issue, we extended the PBSD with sentences from the 350 Android privacy policies (APP-350)~\cite{Sebastian_Sebastian_Zimmeck_Peter} only for permissions that map with Alexa's data permission categories, i.e.: the \emph{Device Address}, \emph{Device country and postal code}, \emph{Email Address}, \emph{Location Services}, and \emph{Mobile Number}. 

\begin{table}[ht]
\vspace{-2.4mm}
\centering
\footnotesize
\caption{Sentence count per %
permission %
in PBSD and the extended PBSD+APP350.}
\label{tb:sentencecount}
\begin{tabular}{r|r|r}
Alexa Data Permission           & \#PBSD & \#PBSD+APP350\\
\hline
Amazon Pay                     & 135    & 135          \\
Device Address                 & 254    & 776          \\
Device Country and Postal Code & 244    & 518          \\
Email Address                  & 205    & 2181         \\
Location Services              & 186    & 533          \\
Mobile Number                  & 97     & 941          \\
Name                           & 416    & 416          \\
Personal Information           & 2,482  & 2,482        \\
None                           & 6,390  & 6,390        \\ \hline
Total                          & 10,409 & 14,372       \\
\end{tabular}
\vspace{-3.2mm}
\end{table}

The problem we tackle is a multi-label classification problem, as sentences can be referring to more than one permission. For instance, the sentence ``We collect your name and postal address'' indicates the collection of permissions \emph{Name} and \emph{Device country and postal code}. 
One well-known way to tackle multi-label classification problems is to transform the problem into a number of binary classification problems.\footnote{Note that other transformations may be possible~\cite{madjarov2012extensive}, e.g., transforming the multi-label classification problem into a multi-class classification problem but that would require $2^8$ classes.} 
In particular, the sentence classifier is formed by 9 binary models, one per each of the 8 Alexa data permission categories we model, and one for the sentences that do not belong to any permission category. 
All binary classifiers are trained following a one-vs-all strategy. 
For every classifier modeling a particular permission, we label all sentences relating to that permission as the positive class and all other sentences as the negative one. 
Given a sentence from a privacy policy, each of the classifiers then evaluates whether that sentence belongs to the permission category being tested or not. Therefore, we obtain a multi-label classification of the sentence. 
That is, every classifier makes an independent assessment and as a result we are able to model sentences describing more than one permission. 

\subsection{Traceability Analyzer}\label{sec:ta}
The traceability analyzer collects the data permissions outputed by the Sentence Classifier, and compares them with the permissions requested by the skill in two steps:

\noindent\textbf{Policy Preprocessing.}
Given an Alexa policy, the Traceability Analyzer first splits the policy into sentences and preprocesses each of the sentences in the same way as presented in \S\ref{sec:pt}. 
After this, it removes sentences non-related to data permissions, such as those in which the authors provide some contact details, e.g.: ``if you have any inquiries about this skill, please contact us by email''. 
These sentences are often structured similarly in all policies, use similar terms, and are usually wrongly assessed by classifiers, which are unable to identify that the sentence does not describe a data permission request. 
To detect this special set of sentences, we use a keyword blacklist containing words such as `contact us' or `call us'. The same is done with negation terms such as ``does not'' or ``doesn't''.
This step filters out negative sentences such as ``this skill does not need your email address'', which are quite common among skill policies. If a sentence contains any of these blacklisted keywords, the sentence is ignored and not classified, as
modeling negative statements is  important to determine the correct meaning of a privacy policy~\cite{andow2019policylint}. 

\noindent\textbf{Traceability.}
After the policy preprocessing, the Traceability Analyzer runs each of the 9 classifiers through all the remaining sentences, collecting all data permissions in the policy. 
Note that the None classifier prevails over the others to handle contradictions in data policies~\cite{andow2019policylint}. 
That is, we consider that a policy does not have a (proper) data statement if there are permissions associated to a policy in addition to a positive classification of the None classifier. 
Next, through the comparison between permissions requested by the skill and those automatically found in its policy, the Traceability Analyzer classifies the traceability of the skill as \emph{broken}, \emph{partial} or \emph{complete}, in the same way as in \S\ref{sec:permissions}.

\subsection{Implementation and Evaluation}\label{sec:syseval}

\noindent\textbf{Sentence Classifier Training and Validation.}
Each of the 9 models is a binary SVM\footnote{ {SVM is suited for binary %
classification problems \cite{cortes_vapnik_1995}, as we do not have enough data to use deep learning techniques, and it usually performs better for NLP tasks with n-grams than other classifiers~\cite{aiyar2018n}.}} classifier built on top of an n-gram binary vectorizer and a tf-idf layer\footnote{Note that approaches such as sentence embeddings \cite{pagliardini2017unsupervised} could be used, but the repetition of similar constructs among policies indicates that a simpler, n-gram classifier is also appropriate}.
We test various parameters over a set of 5-fold cross-validation runs for each of the 9 models, including the size of the n-grams, the SVM loss method, the SVM alpha value, and different oversampling and undersampling strategies to balance the classes of the \emph{permission-by-sentence} dataset (PBSD).
We make sure that each data permission is represented consistently among the instances selected for training and testing in order to increase the robustness of the classifiers against sentences of all types.
The parameters that consistently returned better F1 and accuracy scores over the different 5-fold cross-validation are then used to train with all data. %
In a nutshell, the best parameters are: 
we use n-grams of size 1, 2 and 3, stratified random undersampling and oversampling as balancing strategy, \emph{modified huber loss} as the SVM loss function, and an \emph{svm alpha} of $10^{-5}$ for the SVM classifier. 
After undersampling and oversampling certain classes, on average, each classifier was trained and tested using between 2K and 8K of the total sentences found in the \emph{permission-by-sentence} dataset. %
Interestingly, we observe that the classifiers performs best when we use {\em at the same time} n-grams of different sizes (1,2,3). 

\begin{table}[t]
\caption{K-fold validation %
for the Sentence Classifier.} 
\vspace{-3pt}
\label{tb:f1scores}
\centering
\footnotesize
\begin{tabular}{|l|rr|}
Alexa Data Permission          & F1                    & Acc            \\
\hline
Amazon Pay                     &  .987                 &  .991          \\
Device Address                 &  .919                 &  .908          \\
Device Country + P.C.          &  .921                 &  .909          \\
Email Address                  &  .967                 &  .965          \\
Location Services              &  .960                 &  .973          \\
Mobile Number                  &  .978                 &  .977          \\
Name                           &  .977                 &  .983          \\
Personal Information           &  .986                 &  .982          \\
None                           &  .993                 &  .991          \\
\end{tabular}
\vspace{-2pt}
\end{table}

The final version of all binary classifiers that are deployed in \system{}, are trained using 100\% of the the PBSD dataset, and then evaluated using a set of extra \emph{unseen} 523 sentences to check their performance at a sentence-permission level (see below for the evaluation of \system{} as a whole).
The average F1-score and accuracy metrics obtained %
are reported in Table \ref{tb:f1scores}. 
Note how the sentence classifiers are able to differentiate and classify each of the permissions correctly, obtaining F1-scores and an accuracy of over .9 for all data permissions. 

\noindent\textbf{SkillVet Evaluation.}
To evaluate the performance of \system{} as a whole, we compare the traceability results we obtained with the \emph{unseen} remaining subset of 972 skills and their policies from the \emph{traceability-by-policy dataset (TBPD)}\footnote{Note that we did not consider a further 254 skills in TBPD, because their traceability is trivial: they have missing policies, dead links, or empty policies. A simple check without sentence classification is enough to assess them broken. Including them would potentially bias the evaluation away from the more complex, error-prone cases.%
}. 
That is, the policies of these skills are not used to create the PBSD as detailed in~\S\ref{sec:sc}, so they are not used at any point for training, validation, fine-tuning, or evaluation of the sentence classifiers. 
{It takes \system{} 7 mins 23 secs to process all 972 skills (an average of 0.45 secs per skill).}
As shown in Table~\ref{tb:traceablitycomparison}, \system{} is able to correctly classify 93.1\% (905) %
of the previously unseen 972 policies. The highest accuracy measures are obtained when identifying broken and complete policies (98.5\% and 93.9\% accuracy), while partial traceability seems to work comparatively worse, with a few miss-classified as complete.

\begin{table}[t]
\centering
\caption{Confusion matrix comparing \system{} with a human analysis for the 972 \emph{unseen} skills from TBPD.}
\label{tb:traceablitycomparison}
\begin{tabular}{r|lll|l}
\multirow{2}{*}{\system{}} & \multicolumn{4}{c}{Actual Traceability on 972 unseen policies}  \\
 & Broken (B)  & Partial (P) & Complete (C) & Total \\ 
\hline
B                                                                                 & 264 (98.5\%) & 8  (4.9\%)   & 11   (2.0\%)  & 283                    \\
P                                                                                 & 0   (0.0\%)  & 134 (81.7\%) & 22   (4.1\%)  & 156                    \\
C                                                                                 & 4   (1.5\%)  & 22  (13.4\%) & 507  (93.9\%) & 533                    \\ 
\hline
Total                                                                             & 268         & 164         & 540          & 972  \\              
\end{tabular}
\vspace{-3.0mm}
\end{table}

We further investigate the reasons behind the 67 errors in total (out of 972 skills), which we organize into the following three main types: i) \emph{Class error} (32 cases, 48\%) occurs when the policy is wrongly analyzed due to an error in one of the sentence classifiers producing the wrong class; ii) \emph{Fair error} (15 cases, 22\%) occurs when even a human struggles to identify the correct class because the sentence can be interpreted in different ways, e.g., the most common cases are sentences that refer to {\em address} but it is not clear whether it is the device address or the email address, in fact, the manual analysis for traceability would consider this as partial precisely because it is not exactly clear what address the sentence is referring to; iii) \emph{Filter error} (20 cases, 30\%) is when the error is attributed to the preprocessing stage of \system{}, e.g., most of the cases are due to sentences where there is a complex contradiction, in the sense that there is a negation but associated to a positive non-collection action, actually suggesting a good practice (e.g., not shared with, not disclosed to) in an overall positive sentence.

We next discuss how one might be able to improve the performance of \system{} further in light of the errors. 
For Class errors, one way to improve would be to have more tagged sentences in the PBSD. In general, however, \system{} shows a very good performance, considering that PBSD has less than 250 instances for most of the permissions (cf. Table~\ref{tb:sentencecount}). 
For Fair errors, this is more challenging, but one possible improvement could consider a separate sentence classifier trained with cases where it is not clear what address is being considered, where a human looking at the wider context can make the right assessment (though in most cases as stated above, the human tagger may not be able to ascertain this either). 
Finally, for Filter errors, there may be room for improvement by considering a more sophisticated preprocessing approach perhaps aided with an ontology, similarly to~\cite{andow2019policylint}. However, this may also introduce errors in turn, and the way \system{} accounts for negations and contradictions seem effective in general for this domain. %

\section{Beyond API Permissions}
\label{sec:bypassing}
Amazon enforces access to personal data through a permission model embedded in their APIs as discussed in \S\ref{subsec:data-permissions}. 
However, skills could also request personal data directly from the user without Amazon's API.
This can be done via the Web through account linking, or via conversations. %
\subsection{Account Linking}
\label{Account_Linking}

Account linking allows users to connect their identity with the one they use in a different system like Google, Amazon, or Facebook~\cite{alexa_developer_account_linking}. This is implemented in Alexa using OAuth 2.0.
A total of 5,230 skills (about 4.7\% of the skills in our dataset) are using the account linking feature across marketplaces (with the  breakdown: FR-23\%, DE-10\%, US-6\%, UK-6\%, AU-4\%, CA-4\%).
This is over twice the 2,608 skills (2.3\%) that request for at least one of the Alexa permissions (see beginning of \S\ref{sec:permissions}).
By systematically enabling skills with account linking using a script developed with Selenium~\cite{webdriver2017selenium}, we report where they connect to. %
To do this, we check whether the skill name or developer's name is in the domain used on the account linking URL, and/or whether a name of an OAuth service provider is present (for curation process and full list of OAuth providers refer to \ifthenelse{\boolean{longver}}{Appendix~\ref{sec:account_linking}}{\cite{edu2021skillvet}}).

Table~\ref{tb:account_breakdown} shows that most of the skills connect to the developer's system, although an important number of skills rely on third-party services such as Google or Facebook for authentication. 
In particular, 52\% of the skills with account linking connect to the developer's site, and 29\% to services like Facebook, Twitter, Amazon, or Google, among others. 
Some skills (14\% of all 5,230) allow both,
authenticating through a third-party using OAuth before being redirected to the developer's site. 
Finally, the rest 5\% (Unresolved) could not be labelled as developer or third-party. 

\begin{table}[ht]
\centering
\caption{How skills connect user's Alexa identity with their identity in another system.}
    \label{tb:account_breakdown}
    \vspace{-0.2cm}
\begin{tabular}{m{3.5cm}|m{2.5cm}|c}
\textbf{Account Type} & \textbf{Number of Skills} & \textbf{Percent} \\ \hline
Developer & 2,720 & 52\% \\
Third-Party & 1,517 & 29\% \\
Third-Party or Developer & 732 & 14\% \\
Unresolved & 262 & 5\% \\\hline
\multicolumn{1}{r|}{Total} & 5,230 & 100\%
\end{tabular}%
\end{table}

We then look at the traceability of the skills with account linking and at least one permission. Table~\ref{tb:Account_Traceability} shows that 47\% of them have broken or partial traceability. %
While these broken and partial results are conclusive, when it comes to \emph{complete} traceability, results may not be conclusive --- recall that account linking could enable the collection of further personal data without the need to use data permissions (i.e., taking the user out of Alexa). 
Accounting for the exact data collected by developers through account linking is challenging because: 1) third-party sites used for account linking have all different formats and are thus difficult to scrape; 2) developers may ask for any type of personal information (at any point, and not just during the registration).

\begin{table}[ht]
\centering
\renewcommand{\arraystretch}{0.9}
\caption{Traceability results for English-speaking Markets skills with account linking and at least one permission.}
\label{tb:Account_Traceability}
\scalebox{0.9}{
\begin{tabular}{l|l|l|l|l|l|l|l}
                                 & \multirow{2}{*}{Skills} & \multicolumn{2}{c|}{Broken} & \multicolumn{2}{l|}{Partial} & \multicolumn{2}{c}{Complete} \\ \cline{3-8} 
Market                           &                         & N           & \%            & N            & \%            & N             & \%            \\
\hline
US         & 133        &  30         & 22.6\% & 35         & 26.3\% & 68         & 51.1\% \\
UK         & 46         &  6          & 13.0\% & 13         & 28.3\% & 27         & 58.7\% \\
CA         & 30         &  1          & 3.3\% & 6          & 20.0\% & 23         & 76.7\% \\
AU         & 27         &  2          & 7.4\% & 4          & 14.8\% & 21         & 77.8\% \\
IN         & 26         &  5          & 19.2\% & 5          & 19.2\% & 16         & 61.5\% \\
 \hline
\multicolumn{1}{r|}{Total}  & 262                     & 44          & 16.8\%          & 63           & 24.0\%          & 155           & 59.2\%          \\ 
\multicolumn{1}{r|}{Unique} & 164        &  39         & 23.8\% & 38         & 23.2\% & 87         & 53.1\% \\ 
\end{tabular}}
\end{table}

{To study further the process of account linking and data involved, we randomly choose 70 skills and perform the account linking process manually. 
We see 65 skills (92.85\%) requesting personal data without using the Alexa API. %
We next discuss some case studies we find as a result of this analysis. 
First, we look at ``GCSE Revision'', a skill developed by \emph{Palm UK Ltd} that helps users to prepare for exam-board specific GCSE (General Certificate of Secondary Education) science questions. 
While the skill asks for permissions {Full Name} and {Email Address}, we see that during the linking process the skill asks, in addition, for the user's \emph{address}. 
This is particularly problematic as the skill is   {effectively bypassing Alexa's permission system}, %
using account linking as a proxy to obtain additional data from users.
This also shows a prevalent issue we have discussed earlier (see \S\ref{subsec:traceability_results}):
this skill targets students between {12} and {16} year old students. 
Since GCSE has access to the Full Name, the skill could have been collecting the family name of children in the UK that are preparing for the test, together with how well they perform.  
Note that children's last names can be usually inferred even if the Amazon account was under their parent's names. 
Therefore, the developers of GCSE could send targeted postal advertisement to the household. 
Finally, we also look at ``DaddySays'' skill by \emph{Tagrem Corp} in the Kids category. 
The skill collects the email address of the users during the account linking process in addition to the information collected through the permissions Lists Read Access and Lists Write Access. }

\subsection{Collection via Conversation}
\label{Interaction}

We interact with 100 randomly-chosen skills from those that do not request permissions using the Amazon API and that do not use the account linking feature either. %
We see that 35 are single-interaction skills (they just provide an answer to the user without further interaction), 45 are conversational skills, and the remaining 20 return no response or responded with error (e.g., ``Sorry; I am not sure about that''). 
Out of the 45 conversational skills, we find \emph{3 skills} asking for personal information directly when conversing with users, effectively bypassing the Amazon model for data collection practices. These are `Would You Rather for Family'' by \emph{Voice Games}, ``The Bartender'' by \emph{Pylon ai}, and ``Phone Tracker'' by \emph{S4 Technology, Inc}. For instance, the latter asks for a phone number as part of the conversation. %

To study conversational skills more systematically, we used our system, \system{}, in tandem with recent work, SkillExplorer~\cite{255322}, which has provided an initial method to discover these skills. In particular, we studied the dataset of 100 skills (developed by 89 distinct developers) collecting personal information via conversation provided by the authors of SkillExplorer~\cite{255322}.
Out of all these skills, 82 collect personal information only through conversation, 11 collect personal information via Alexa API in addition to the one collected during a conversation, and 7 also use account linking.
Table~\ref{tb:PII_converse} summarizes the collection method used and shows the results of the traceability analysis. 
Overall, we see that 35\% the skills exhibit broken traceability. 
In this case, all issues are attributed to personal data collected alone during a conversation. 
We also see that 20\% partially disclose their data practices in their privacy policy, and the remaining 45\% exhibit complete traceability. 
One good example of a broken skill is the ``Praise Me'' skill by \emph{Jackson Jacob} with skill id B07G5B4P7R and a customer rating of 4.8 out of 5 in the US market. This skill is available across all five English speaking countries. 
According to its description, this skill is supposed to praise the user after asking for their name. 
However, the skill not only request the name of the user during the conversation without declaring the permission for it, but it also fails to disclose its data practices in a privacy policy. 
Furthermore, the skill has a privacy policy link that takes the user to a porn site. 
There are also two developers with skills having different traceability outputs. 
\emph{Voice first tech} has skills that exhibit partial and complete traceability (1 skill in each category), while \emph{Volley inc.} has broken and complete traceability skills (1 broken, 2 complete skills).

We also compare the results given by \system{} with a human-based traceability analysis made by one of the authors for the 100 skills. %
As shown in Table~\ref{tb:evalconver}, \system{} achieves a very good accuracy, being able to correctly identify the traceability for 90 of the 100 skills. It is particularly very good at spotting complete (95.2\%) and broken (94.3\%) traceability, and good for partial (73.9\%). Of the 10 errors in total, 8 are Class errors, 1 Filter error and 1 Fair error. These results align with what we report for the unseen 972 skills that request permissions through Alexa API in \S~\ref{sec:syseval}. 

Finally, note that SkillExplorer is restricted by design to those skills that have unique invocation names~\cite{255322}. Future work should look into methods to discover this type of skills %
considering skills that share invocation names, which are very prevalent, as shown in this paper (\S~\ref{sec:overview}). 
Importantly, however, \system{} would easily integrate with any such tool when provided with the data the skills ask for and any accompanying privacy policies.

\begin{table}
\centering
\caption{Traceability Result with \system{} on Subset of skills found in~\cite{255322} collecting personal information via conversation.}
\label{tb:PII_converse}
\begin{tabular}{|l|l|l|l|l|l|} 
\hline
\multicolumn{1}{|c|}{\multirow{2}{*}{How PII were collected}} & \multicolumn{3}{l|}{Traceability} & \multicolumn{2}{c|}{Total}                   \\ 
\cline{2-6}
\multicolumn{1}{|c|}{}                                        & B  & P  & C                       & Skills               & Dev                   \\ 
\hline
\hline
Conversation + Account Linking               &    & 1  & 6                       & 7                    & 5                     \\ 
\hline
Conversation + Alexa API                                      &    & 5  & 6                       & 11                   & 9                     \\ 
\hline
Conversation Only                                             & 35 & 14 & 33                      & 82                   & 79                    \\ 
\hline
\hline
\multicolumn{1}{|r|}{Total number of \bf Skills}                                   & 35 & 20 & 45                      & 100                  & --                    \\ 
\hline
\multicolumn{1}{|r|}{Total number of \bf Developers}                              & 35 & 16 & 40                      & \multicolumn{1}{l|}{--} & \multicolumn{1}{l|}{89}  \\
\hline
\end{tabular}
\end{table}

\begin{table}
\centering
\caption{Confusion matrix comparing \system{} with a human-centered analysis for the set of the skills given by~\cite{255322}.}
\label{tb:evalconver}
\vspace{-2.4mm}
\begin{tabular}{r|rrr|l}
\multirow{2}{*}{\begin{tabular}[c]{@{}r@{}} \system{} \end{tabular}} & \multicolumn{4}{c}{Actual Traceability}        \\
                                                                                  & Broken (B) & Partial (P) & Complete (C) & Total  \\ 
\hline
B                                                                                 & 33 (94.3\%) & 2 (8.7\%)    & 0 (0\%)       & 35     \\
P                                                                                 & 1 (2.9\%)   & 17 (73.9\%)  & 2(4.8\%)      & 20     \\
C                                                                                 & 1 (2.9\%)   & 4 (17.4\%)   & 40 (95.2\%)   & 45     \\ 
\hline
Total                                                                             & 35         & 23          & 42           & 100   
\end{tabular}
\end{table}

\section{Discussion}
\label{sec:DISCUSSION}

\subsection{Key Findings and Limitations}
\noindent\textbf{Potential privacy violations}.   
We see bad privacy practices in about 748 skills ({43\%}) of those that request permissions in English-speaking marketplaces\footnote{Recall English-speaking skills are 82.7\% of the total}%
, involving 557 ({50\%}) of the developers with skills that request permissions. Although it could be some developers have multiple Amazon developer accounts and the real number of entities behind those accounts are less, this still paints a very worrying picture. %
{Particularly worrying cases are those in categories like Kids, Schools, and Education, which exhibit broken (72 skills in total) and partial ( 17 skills in total) traceability.}
Beyond the traceability itself, this may have additional implications, as skills in those categories might be subject to even tighter privacy regulations for children, such as the US Children's Online Privacy Protection Act (COPPA) of 1998 \cite{Coppa_1998}. Other categories, such as the {Utilities} category, also have countless examples of unjustified collection of personal information {-- i.e., with broken traceability}. 
Skills, such as ``Mock Interview'' \emph{by Graylogic Technologies}, ask for permissions --- like Device Address --- without acknowledging such collections in its privacy document.  
Furthermore, we even see a case of a skill collecting information, while stating that it is not currently used. This is the case of the ``Bin reminder'' skill by \emph{Shane} that collects the user's device address. 

\noindent{\bf Bulk skill creation}. 
There are a few developers that publish hundreds and even thousands of skills (cf. \S\ref{sec:overview}). 
We see some developers using automated systems to develop and/or deploy skills. 
This democratizes the creation of skills, but also lowers down the entry barriers for criminals and can foster the commoditization of unwanted skills. 
Previous works have recently shown that miscreants leverage online app generators to publish unwanted apps in Android~\cite{kotzias2021did}. 
We believe that Amazon Alexa will suffer from the same challenges than Google or Apple with the vetting of apps in Android or iOS markets~\cite{Hu2020SecurityVP}. {In fact, recent research seems to point towards issues with the current vetting process in Alexa~\cite{chengdangerous2020}.} %
We posit that the research community needs to develop detection mechanisms tailored to the SPA ecosystem. 
However, one important limitation is the lack of access to the skill software (e.g. deployed in the cloud), which also constrains our analysis. 

\ifthenelse{\boolean{longver}}{
\noindent{\bf Potential for impersonation and Sybil attacks}. 
We see both thousands of skills from different developers with a similar name (impersonation), and hundreds of skills from the same developers in the same market (sybils).
We have seen how this may pose a threat to the invocation system of SPAs.  
Alexa has recently introduced a mechanism to provide name-free skill interaction, i.e., 
a skill can be invoked without its skill invocation name. 
For this, Amazon uses rich contextual information to select the best skill that matches the user request~\cite{DBLP:journals/corr/abs-1804-08064}. 
This widens the attack surface as impersonation attacks may not necessarily need to craft skills with the exact same name. 
While our work does not consider name-free skill in our impersonation analysis, our findings can be seen as an under-approximation of the problem, suggesting that motivated attackers could take advantage of this.}

\noindent{\bf Market migration}. 
Amazon allows developers to publish the same skill in different markets. 
Our study shows that there is an important overlap of skills across markets, which includes the use of the exact same privacy policy. 
This has important regulatory implications derived from the different local regulations in a globalized ecosystem, similar to what we have seen for cookies in the wake of GDPR~\cite{Xuehui20_eurosp, IMC2018-GDPR}, and the different languages across marketplaces. %
Note, however, that skills that are pushed from English-speaking markets to other markets now go through a language migration process, and privacy policies may differ for the same skill~\cite{alexa_English_Variant}.  
This process was not in place 
{at the time we started our study.} 
As future work, we plan to re-crawl all 
{policies} and study how many have changed with the skill migration feature. 
Another limitation is that we only conduct the traceability analysis for English-speaking marketplaces,  %
as this is the source language of our ground-truth. 
We could overcome this limitation using automated translation, but this could introduce errors in the classification. 
Evaluating how our model performs with translated policies is part of our future work.

\ifthenelse{\boolean{longver}}{
\subsection{Other Concerning Practices}
\label{subsec:Concerning_Practices}

\noindent\textbf{Advertisement:}
We see evidence of skills embedding advertisement as part of their responses. A good example is the ``myTuner Radio Player Canada'' skill by \emph{Appgeneration Software technologies} where a full-screen ad pops up (in SPA device with a screen) when user select a different station. %
Another example is the ``Sleep and Relaxation Sounds'' skill by \emph{Voice Apps, LLC} which keeps advertising its premium subscription and overwhelms users with constant commercials until they buy its premium service.

\noindent\textbf{Spamming:}
We observe several skills spamming users for reviews, like ``Sleep Sounds: Ocean Sounds'' by \emph{Invoked Apps LLC} and ``Good Morning Gorgeous'' by \emph{Skillex Studios}. 
Also, there are skills such as ``Hits 1 Latina'' by \emph{autopo.st}, and ``Sleep and Relaxation Sounds'' by \emph{Voice Apps, LLC} that do not respond to \emph{Alexa stop command} and keep spamming users with unsolicited information after invocation. ``Night Light'' by \emph{labworks.io ltd}, requires users to say \emph{``Alexa stop nightlight''} instead of the usual stop command.
``Sleep Sounds: Pink Noise'' by \emph{Voice Apps, LLC} both spams users for a review and fails to respond to Alexa stop command.
}

\subsection{Developers' Business Models}
\label{sec:motivations}
Our key findings warrant a further discussion about the motivations developers may have to develop skills. In this regard, our study identifies over {47K developers that have contributed to the Alexa marketplace with a rich ecosystem of skills (cf. Table \ref{tb:Skills and Developers} in \S\ref{sec:overview}). 
Since Amazon forbids advertisements in skills, an important open question is what do these developers gain and what are their motivations.} First, companies may be interested in offering a voice-over interaction with the user through Alexa. 
Examples are skills in the Travel \& Transportation category (e.g. to order a ride) or in the Food \& Drink category (e.g. to order a pizza).
In these situations, developers need to bind the user's Amazon account with the external service via account linking, as mentioned before.
Interestingly, we only {observe 5\% of the skills requesting account linking, which is circa 5K unique skills (cf. \S\ref{Account_Linking}). %
We can, thus, conclude that account linking is not yet the primary motivation for most developers.}

{Amazon promotes in-skill purchasing and paid subscriptions, %
which allows developers to sell premium content in skills such as extra features, in-game elements and interactive stories. 
This skill content can be offered as: i) subscriptions, where developers charge users recurrently to access premium features, ii) one-time purchases, where users pay once to have permanent access to the premium features and content, and iii) consumables, where content can be purchased, exhausted and purchased again (e.g., extra-lives in games~\cite{alexa_In_Skill_Purchases}).
In-skill purchasing %
is not yet supported by all markets. 
At crawling time, we only see 1,535 (1.37\%) skills with in-skill purchasing or paid subscriptions in the US and UK. 
Prices for in-skill products range considerably, from 0.99 to 99 USD/GBP.  %
``Sleep Sounds: Hair Dryer'', ``Sleep Sounds: Harp Sounds'', and `` Sleep Sounds: Heavy Rain'' developed by \emph{Voice Apps, LLC.} are examples supporting in-skill purchases.
However, judging by the number of skills with  in-skill purchasing, we can conclude that this is not the main source of income yet.}
Developers may also offer paid skills in the future, but this is not an option at writing time.

Overall, we see that most of the skills in our dataset are free (both to enable and to use fully). 
The total number of skills that are free and do not have in-skill purchasing is 110,494 (98.6\%). 
Out of these, 106,289 (94.9\%) do not offer account linking either. 
Therefore, though there might be small incentives, such as the Amazon incentive program~\cite{Amazon_incentive_program}, which rewards skill creation (e.g. with a smart plug), or the simple curiosity to develop in a new environment, %
we conclude that {\em there is an ample number of skill developers for which there is no clear business model}. 
Some of these skills might be in the data monetization business.
An obvious example is ``Pixated Salat'', the prayer skill developed by a large advertisement company (c.f., \S\ref{subsec:traceability_results}). 
Another example is the ``Autochartist'' skill, %
which states in their privacy policy that the data they collect may be used {``to provide you with news, special offers and general information about other goods, services and events which we offer that are similar to those that you have already purchased or enquired about unless you have opted not to receive such information''}. 
It is, however, unclear how users of skills can opt out.

\subsection{Mitigation and Responsible Disclosure}
Tackling bad practices from third-party developers is a challenge to systems security. 
Having a well-defined --- fine-grained --- permission model is an important step forward. 
However, we have seen that: 1) developers can bypass it, and 2) developers do not offer transparency about how they utilize users' information. 
We next present a range of countermeasures that could mitigate the risk of enabling third-party developers access to users' data. 
First, Amazon should not allow developers to reuse policies verbatim as this is error-prone, as we have seen. 
Second, both Amazon and the developers should perform a thorough review of the traceability themselves. 
\system{} can help in the process of systematically reviewing permissions and privacy policies. 
However, further research efforts are needed to ensure that policy notices are relevant, actionable and understandable by users~\cite{schaub}. 
Third, publishing the code of the skills or the binaries will considerably foster research in the area, similar to what we have witnessed in Android~\cite{chen2015finding}. %
While approaches to detect unwanted skills can be adopted from Android, the Alexa ecosystem has unique features like voice recognition that present novel challenges (e.g., voice masquerading attack~\cite{zhang2019dangerous} --- malicious skills that pretend to hand over the control to another skill). 
Finally, another important mitigation is to increase user awareness, %
as users are known to have incorrect or at best incomplete mental models of how assistants such as Alexa work~\cite{238321, Noura_Abdi_2019}. 
This may be even more challenging when users interact with skills aside from the policy enforcer, such as during the account linking process or with conversational skills.  
There have been works to defend users against malicious activities on online services (e.g., spam~\cite{stringhini2010detecting}, or phishing detection~\cite{fette2007learning, dou2017systematization}). 
Solutions in this direction will have to consider that the assistants' nature brings, again, unique challenges.
We refer the reader to~\cite{DBLP:journals/corr/abs-1903-05593} for a security and privacy review of countermeasures for Alexa-like assistants.  

\noindent\textbf{Responsible Disclosure:} 
We perform a responsible disclosure process, starting from mid August 2020, as follows.
First, we notify all skill developers who are not engaging in good data practices whenever we have their contact details.  
Second, we also report our findings to Amazon and have confirmed that the skill store team have taken action. 

\vspace{.2cm}
\noindent \fbox{\parbox{0.97\columnwidth}{
As a result, we have already seen that 150 skills --- 13\% of which we identified issues for --- no longer pose a threat to users at the time of writing: 137 of these skills have been vetted and are no longer available on Alexa, while 13 of them now have complete traceability. }}

\section{Related Work}
\label{sec:Related Study}

\noindent{\bf Skill Measurement, Privacy and Security.}
There have only been limited related work on SPA third-party skills and the mechanisms they use in the discovery process \cite{white2018skill}.
There has been work on measuring the availability of privacy policies for skills \cite{alhadlaq1902privacy}, and on whether the vetting process in Alexa can be compromised~\cite{chengdangerous2020} (by crafting 234 mock policy-violating skills).
However, previous works do not focus on data practices by actual skills developers and do not perform a traceability analysis between privacy policies and data permissions asked by the skills. %
The authors in \cite{liao2020measuring} used skills description as a baseline to detect inconsistent privacy policies, but not the actual data permissions collected by the skills through Alexa API. This is a limitation as many developers don't often mention what permissions they collect when describing their skills. For example, skills developed by \emph{Vipology} such as ``100.7 BIG'', ``FM101.7'', ``KISS FM 101.7'', %
and ``The Wolf 103.3'' request for Device Country and Postal Code through Alexa API. However, Vipology did not mention this permission in its voice-app descriptions. %
Besides, the permissions mentioned by the developers in the skill descriptions could be different from what the skills are requesting for. A good example is the ``Interflora`` by \emph{Interflora British Unit}. This skill collects ``Full Name'', ``Email Address'', ``Mobile Number'', and ``Amazon Pay'' permissions through Alexa API but only mention the Amazon pay permission in its skill description. 

Authors in \cite{255322} conducted a measurement study to understand skills’ behaviour by building a system called SkillExplorer. This system interacts with skills and identifies those that request personal information through conversations bypassing developer specifications. However, the system only works with a limited number of skills: those with a unique invocation name, so they missed thousands of skills with similar or same invocation name --- cf. Section~\ref{subsec:skills-developers}.
Lentzsch et al.~\cite{lentzschhey}, in parallel to our work, %
 provided an initial look at the effectiveness of skill privacy policies. We, however, offer a much more nuanced view of the data practices in third-party skills and their developers. This allows us to uncover many more partial traceability cases. Our crawling strategy also allows us to analyze the traceability of 1,758 unique skills (56\% more skills). %
Together, this brings the number of skills with complete traceability to 57\% (instead of the 77\% reported in~\cite{lentzschhey}). Besides, their automated method {that uses PoliCheck~\cite{andow2020actions} underneath} has 83.3\% precision when detecting complete policies, whereas \system{} has a precision of 95\%. This better precision could be attributed, among others, to \system{} covering all data types in Alexa (incl. Lists and Amazon Pay), while in~\cite{lentzschhey} the ontology they use for PoliCheck only considers a subset of the data types. Further, PoliCheck was trained \emph{only} with android privacy policies, while \system{} was trained with Alexa skill privacy policies, only complemented with a few android privacy policies (c.f \ref{sec:sc}).

Research on skills also focused on exploiting the speech recognition capabilities of SPA by crafting malicious skill invocation names, usually leveraging phonetic similarity~\cite{217575,zhang2019dangerous}. 
More recently, a study in~\cite{Security_Research_Labs} shown how malicious developers could potentially sneak malicious code into their software via the application backend after vetting, and another interesting study looked into how a malicious third-party skill could covertly reword responses from legitimate sources to intentionally introduce misperception about the reported events \cite{SHAREVSKI2021102604}. There is also a report of how Amazon bug could allow an attacker to access a user's install Alexa skills and user's Amazon profile information through a phishing attack~\cite{Amazon_security_bug}. 
Our threat model in this paper is different from these works (see Section \ref{sec:threat}).

\noindent{\bf Privacy Traceability Analysis.} 
Our traceability analysis relates to previous work that focuses on ensuring software requirements
compliance with governing legal texts and privacy policies \cite{young2010method}. In particular, previous work analyze privacy traceability in domains such as Online Social Networks \cite{anthonysamy2013social}%
Social Media Aggregators \cite{misra2017privacy}, and Smartphone Applications \cite{Sebastian_Sebastian_Zimmeck_Peter}. %
We, however, focus in this work on analyzing the privacy traceability of skills. Also, in \cite{Sebastian_Sebastian_Zimmeck_Peter}, the authors analyze Android apps characteristics for likely non-compliance with privacy requirements and some selected and applicable laws. In our case, we only analyze non-conformity with privacy requirements based on the requested data permissions by the skills.
Finally, PolicyLint~\cite{andow2019policylint}, a tool to identify policy contradictions in Android, is complementary to our work, and it could be used to understand if there are ill-defined policies in Alexa.
While self-contradictions are orthogonal to the traceability analysis we do, \system{} accounts for negations in a way proven effective for traceability analysis. 
However, we plan to explore the use of ontologies as in~\cite{andow2019policylint} as future work.

\section{Conclusion}
\label{sec:CONCLUSION}
The Amazon skill ecosystem has grown rapidly without a clear business model. 
We have presented a measurement study that provides a %
large-scale analysis of this ecosystem, analyzing over 199k third-party skills. 
By looking at the distribution of skills and the diversity of developers, we have shown current practices in the wild and offered a unique understanding of developers' motivations.  
As a key novelty, we propose an automated system, \system{}, that analyzes the traceability of permissions. 
While transparency is paramount to let users make informed decisions about disclosing their data, our result shows that 43\% of skills do not comprehensively disclose their data practices.
We have also seen how skills may bypass Alexa's permission system by requesting personal information without using their APIs (e.g., on an external domain via account linking). 
This indicates that even skills with a complete privacy policy can pose a risk to users. 
We have discussed the most concerning practices and the implications of our findings, with our responsible disclosure helping address 13\% of the issues found at the time of writing.

\subsection*{Acknowledgments}
We thank the anonymous reviewers. 
This research was partially funded by EPSRC under grant EP/T026723/1 and the ``Ramon y Cajal'' Fellowship RYC-2020-029401-I. 
Edu was supported by the Petroleum Technology Development Fund (PTDF) for his PhD.

\bibliographystyle{IEEEtran}
\bibliography{reference}

\ifthenelse{\boolean{longver}}{
\appendix

\subsection{Account Linking Details}
\label{sec:account_linking}

\noindent\textbf{List of OAuth Providers.}
\label{OAuth2.0_providers}
To compile the list, we look at 100 skills with account linking features selected at random to note which domains they can connect to. Google, Amazon, Twitter, and Facebook are the only 3rd party OAuth2.0 providers we observed. For completeness, we then include other popular OAuth2.0 providers from the top 50 rank websites on Alexa.com. We are confident that we include most OAuth providers judging by the lower number of unresolved accounts in Section~\ref{Account_Linking}.
The full list of OAuth providers we use %
is:
{\it
Amazon, AOL, Autodesk, Apple, Basecamp, Battle.net, Bitbucket, bitly, Box, Cloud Foundry, Deutsche Telekom, deviantART, Discord, Dropbox,  Facebook, Fitbit, Formstack, Foursquare, GitHub, GitLab, Google, Google App Engine, Huddle,  Imgur, Instagram, Intel Cloud Services, Jive Software, kakao, Keycloak, LinkedIn, Microsoft (Hotmail, Windows Live, Messenger, Active Directory, Xbox) NetIQ Okta OpenAM, ORCID, PayPal, Ping, Identity, Pixiv, Reddit, Salesforce.com, Sina, Weibo, Spotify, Stack Exchange, Strava, Stripe, Twitch, Twitter, Viadeo, Vimeo, VK, WeChat, XING, Yahoo, Yammer, Yandex, Yelp, and Zendesk.}

\noindent\textbf{Account Linking Per Categories.}
Table~\ref{tb:Account Linking_Cat} shows the number of skills with account linking per category (only categories with $>5$ skills displayed). We see that Smart Home is the category with the highest number of skills that use account linking feature. This is natural since it is necessary for smart things like connected car, smart home to connect the user's identity with their identity in the developer's system. There are, however, other categories with many skills with account liking too, such as Music \& Audio and Business \& Finance, as these naturally aim to allow the user to connect to their existing online resources through a voice interface.

\begin{table}
\renewcommand{\arraystretch}{1.0}
\centering
\caption {Skills with Account Linking by category (Englishspeaking markets)} 
\label{tb:Account Linking_Cat}
\scalebox{0.8}{
\begin{tabular}{l|l|l|l|l|l}
\textbf{Category} & \textbf{N} & \textbf{\%} & \textbf{Category} & \textbf{N} & \textbf{\%} \\ \hline
Smart Home & 2212 & 42.29\% & News    & 57 & 1.09 \\
Business  Finance & 384 & 7.34\% & Utilities & 56 & 1.07\% \\
Music  Audio & 341 & 6.52\% & Home Services & 36 & 0.69 \\
                  Uncategorized        & 273 & 5.22\% & Movies  TV  & 23 & 0.44 \\
Lifestyle & 249 & 4.76\% & Novelty  Humor & 23 & 0.44 \\
Productivity & 241 & 4.61\% & Sports & 19 & 0.36 \\
Health  Fitness & 226 & 4.32\% & Organizers  Assistants ~ & 18 & 0.34 \\
Games  Trivia & 179 & 3.42\% & kids & 16 & 0.31 \\
Shopping & 150 & 2.87\% & Public Transportation & 16 & 0.31 \\
Education  Reference & 148 & 2.83\% & Knowledge  Trivia & 13 & 0.25 \\
Food  Drink & 127 & 2.43\% & Streaming Services & 13 & 0.25 \\
Social & 98 & 1.87\% & Local   & 10 & 0.19 \\
Travel  Transportation & 94 & 1.80\% & Calendars  Reminders & 8 & 0.15 \\
Connected Car & 79 & 1.51\% & Navigation  Trip Planners & 7 & 0.13 \\
Weather & 59 & 1.13\% & Cooking  Recipes & 6 & 0.11 \\
\end{tabular}}
\vspace{-0.5mm}
\end{table}

\noindent\textbf{Where account linking connects to (examples observed).} %
Table \ref{Sample_skills_in_India} shows some of the skills in the Indian market with account linking and the type of account they connect to.  We can see that skills like ``GoToMeeting for Alexa'', ``Crypto Genie'', and ``Uber'' can connect the identity of users with their identity on the developer's system. While skill such as ``Bollywood Mania'' can only link the user's identity with a third party system, skills like ``HiCare'' can connect the identity of users with their identity on either the  developer's system  and third party system.

\begin{table}
\centering
\caption {Sample skills and account linking domain (India).} %
\label{Sample_skills_in_India}
\scalebox{0.8}{
\begin{tabular}{|m{5cm}|m{3.95cm}|} \hline
\textbf{Skills} & \textbf{Account Type} \\ \hline
GoToMeeting for Alexa, Crypto Genie, J.P. Morgan, Nurturey maths, AGL, Voice Prototypes, Sayspring, Zomato, JioSaavn, Ola, I'm Driving, Phone Genie, Uber, Vodafone & Developer \\ \hline
paisabazaar, StockInvest, Escape the Room, Brightidea Home, Voice Rewards Me, Commvault, & Amazon \\ \hline
Starfish   Local & Google, Amazon \\ \hline
Bing Bong & Facebook, Google \\ \hline
Trivia Monster, The Dark Citadel & Facebook, Google, Amazon, Developer \\ \hline
Bollywood Mania & Twitter \\ \hline
BBC Good Food, cure.fit, Fitbit, KEYCO air  & Facebook, Google, Developer \\ \hline
HiCare & Facebook, Google, twitter, Developer \\ \hline
\end{tabular}}
\end{table}

}

\end{document}